\documentclass[twocolumn]{autart}    

\usepackage{graphicx}          
\usepackage{amssymb}
\usepackage{graphicx}
\usepackage{amsmath}
\usepackage{mathrsfs}
\usepackage{verbatim}
\usepackage{setspace}
\usepackage{arydshln}
\usepackage{color}

\begin{document}

\begin{frontmatter}
\vspace{-5mm}
\title{Inversion-Based Output Tracking and Unknown Input Reconstruction of  {Square} Discrete-Time Linear Systems} 

\author[conc]{E. Naderi and K. Khorasani}\ead{kash@ece.concordia.ca}

\address[conc]{Department of Electrical and Computer Engineering, Concordia University, Montreal, Quebec, Canada}  

\vspace{-6mm}

\begin{abstract}                          
In this paper, we propose a framework for output tracking control of both minimum phase (MP) and non-minimum phase (NMP) systems {as well as systems with transmission zeros on the unit circle}. Towards this end, we first address the problem of unknown state and input reconstruction of non-minimum phase systems. An unknown input observer (UIO) is designed that accurately reconstructs the minimum phase states of the system. The reconstructed minimum phase states serve as inputs to an FIR filter for a delayed non-minimum phase state reconstruction. It is shown that a quantified upper bound of the reconstruction error exponentially decreases as the estimation delay is increased. Therefore, an almost perfect reconstruction can be achieved by selecting  the delay to be sufficiently large. Our proposed inversion scheme is then applied to solve the output-tracking control problem. {We have also proposed a methodology to handle the output tracking prob!
 lem of systems that have transmission zeros on the unit circle in addition to MP and NMP zeros.} Simulation case studies are also presented that demonstrate the merits and capabilities of our proposed methodologies.
\end{abstract}

\end{frontmatter}

\vspace{-6mm}
\section{Introduction}
Output tracking problems arise in many control applications such as in aerospace and robotics. One possible solution to this problem is an inversion-based approach in which the control input is considered as the output of an inverse system which is stimulated by the actual system desired output. However, this approach is quite challenging due to presence of unstable transmission zeros of the system. Unstable transmission zeros also challenge a stable reconstruction of the unknown system inputs using the known outputs. This can be considered as an equivalent problem to the inversion-based output tracking problem. Both inversion-based output tracking and unknown input reconstruction problems have received extensive attention from the control community researchers.

Inversion of linear systems was first systematically treated by Brocket and Mesarovic \cite{Bro}. The classic works are known as \emph{structure algorithm} \cite{silverman}, Sain \& Massey algorithm \cite{SM}, and the  Moylan algorithm \cite{moylan}. Gillijns \cite{gili} has proposed a general form of the Sain \& Massey algorithm in which certain free parameters are available for adjustment based on the design requirements.  However, this is accomplished under the assumption that the original system does not have any unstable transmission zeros.

The problem of  unknown input reconstruction using inversion schemes has been tackled by more sophisticated methods. Palanthandalam-Madapusi and his colleagues (\cite{chavan2015delayed}, \cite{palanthandalam2007unbiased} and \cite{kirtikar2009delay}) have considered the problem of input reconstruction in several papers, yet, the provided solutions are \textit{only} applicable to minimum-phase (MP) systems.  Xiong and Saif \cite{xiong2003unknown} have proposed an observer for input reconstruction that works under limited cases of non-minimum phase (NMP) systems. Specifically, to those non-minimum phase systems with zero feed-through matix ($D$) and those systems having special disturbance dynamics. The restrictive condition of requiring zero feed-through matrix appears in most works that are related to the input reconstruction problem {\cite{wahls}}. Flouquet and his colleagues \cite{floquet2004sliding} have proposed a sliding mode observer for the input recon!
 struction which is also only valid for minimum phase systems with zero feed-through matrix.\\

{A successful solution to stable inversion of minimum phase and non-minimum phase systems was proposed in Zou and Devasia \cite{zou1999preview}. This solution is extended to discrete-time systems in \cite{george1999stable}. However, the solution requires that the system should have a well-defined relative degree. A geometric solution for stable inversion of linear systems is proposed by Marro and Zattoni \cite{Marro2010815}. The algebraic counterpart of the geometric solution is provided in \cite{Marro201111320}.  Both the above geometric and algebraic solutions do not provide a framework for handling systems having transmission zeros on the unit circle in addition to MP and NMP zeros.\\}

The other problem of inversion-based output tracking has also been the subject of a number of research in the literature. It is well-known that unbiased inversion-based output tracking  is essentially non-causal since it requires the information on the entire trajectory in future that is not a reasonable assumption for many applications. Zou and Devasia (\cite{zou1999preview,zou2004preview,zou2007precision}) have introduced \textit{preview-based stable-inversion} method for continuous-time systems. Basically, this method requires access to a finite window of future data instead of having the entire future trajectory, although the approach results in a \textit{degraded output tracking} error performance. This technique has been significantly improved by the recent work (\cite{zou2009optimal,wang2013b}), however, these works are also developed for continuous-time LTI systems. {However, the method is constrained by restrictive assumptions, such as requiring a we!
 ll-defined relative degree condition}. Several other work using different approaches are available in the literature that are mostly application of a particular method such as the Q-learning \cite{kiumarsi2014reinforcement} or filtered basis functions \cite{duan2015tracking}.

In this study, we first address the inversion-based unknown state and input reconstruction problem. A general unknown input observer is proposed that accurately and independently reconstructs the \textit{minimum phase states} of the system by using only the available system measurements. The \textit{minimum phase states} here refers to $n-p$ states of the overall system, where $n$ denotes the order of the system and $p$ denotes the number of unstable transmission zeros. Next, the estimated minimum phase states are considered as inputs to an FIR filter to reconstruct the $p$ \textit{non-minimum phase states} of the system. The FIR filter estimates the non-minimum phase system states with a time delay of $n+n_d$ steps. It also yields an estimation error which is a function of the to be selected parameter $n_d$. We have explicitly derived subsequently the relationship between the reconstruction error and $n_d$.

Specifically, we have shown that the estimation error is proportional to inverse of the smallest non-minimum phase zero to the power of $n_d$. Hence, if the system does not have any transmission zeros on the unit circle, the estimation error asymptotically decays to zero as $n_d$ is increased. This can therefore ensure that an unbiased input and states estimation can be obtained. For most cases, an $n_d$ equal to four or five times $n$ would yield an almost perfect estimation results for any smooth \underline{or} non-smooth unknown input. For a smooth input, an $n_d$ as small as 2 may suffice.\\

We comprehensively address and discuss the dynamics of the non-minimum phase states and have derived the relationships among the system matrices. By invoking a minor modification, our proposed methodology is extended to solve the inversion-based output tracking control problem. As opposed to a delayed reconstruction, our method now requires data corresponding to $n+n_d$ time steps ahead of the desired trajectory.  As in the previous problem, we have quantified the tracking error characteristics and have shown that an almost perfect tracking is achievable by properly selecting $n_d$ that yields an unbiased state reconstruction that can be achieved as in the first problem.\\

{Finally, we have shown that our proposed methodology for stable inversion of linear systems can be successfully extended to handle the output tracking problem in systems that have transmission zeros on the unit circle in addition to MP and NMP zeros. Our proposed solution introduces a further delay of $n_c$ due to implementation of a controller of order $n_c$. In contrast to unstable transmission zeros, the output tracking error does not exponentially decrease by increasing $n_c$. Instead, the output tracking error depends on the norm of a transfer function which is parameterized by the system and the selected controller parameters. We have further characterized design criteria and have formulated a minimization problem for selection of the controller parameters. }

To summarize, the \underline{main contributions} of this paper can be stated as follows:
{
\begin{enumerate}
\item A methodology for estimation of unknown states and unknown inputs of \textit{both} minimum and non-minimum phase linear discrete-time systems is proposed and developed,
\item  In our proposed methodology, the MP states are partitioned and estimated by using the system measurements. The MP states are \textit{exactly} estimated with a delay of at most equal to the system order; in contrast to the available works in the literature where all states are \textit{approximated} with a delay that depends on the location of the smallest unstable transmission zero,
\item Our proposed solution does not require that the system should have a well-defined relative degree,
\item An algorithm and a simple constructive procedure for designing the inversion-based output tracking control scheme is proposed,
\item We have shown that our proposed solution provides a framework for handling the output tracking problem of systems that have transmission zeros on the unit circle in addition to MP and NMP zeros for the first time in the literature, and finally
\item The accuracy of our proposed input and state estimation scheme as well as the output tracking control performance as a function of  the delay parameter are quantified and investigated. For the case that the system has transmission zeros on the unit circle, the output tracking error is characterized by the norm of a transfer matrix that depends on the system and controller parameters.
\end{enumerate}}
The remainder of this paper is organized as follows. The problem statement and preliminaries are provided in Section \ref{sec: problems}. Section \ref{sec:state-input-reconstruction} is devoted to the problem of developing and designing unknown state and input reconstruction methodologies. The problem of developing an inversion-based output tracking strategy is addressed in Section \ref{sec:output-tracking}. {The extension of the solution to the case where the system has transmission zeros on the unit circle is developed in Section \ref{sec:unitcircle}}.  Finally, several numerical case studies are presented in Section \ref{sec:simulation} to demonstrate and illustrate the capabilities of our proposed methodologies.

\vspace{-4mm}
\section{Problem Statement}\label{sec: problems}
Consider the following deterministic discrete-time linear time-invariant (LTI) system $\mathbf{S}$,
\begin{equation}\label{eq:system}
\mathbf{S}: \left\lbrace\begin{array}{l}
x(k+1)=Ax(k)+Bu(k)\\y(k)=Cx(k)+Du(k)
\end{array}\right.
\end{equation}
where $x \in \mathbb{R}^{n}$, {$u \in \mathbb{R}^m$} and $y \in \mathbb{R}^l$. The quadruple $\Sigma:=(A,B,C,D)$ is assumed to be known \textit{a priori}.  The output measurement $y(k)$ is also assumed to be available, however, \textit{both} the system states $x(k)$ and $u(k)$ are assumed to be unmeasurable. In this paper, we consider the following \textit{two} specific problems.

\textit{\underline{\textbf{Problem 1}}: The system states and the unknown input reconstruction:} The objective of this problem is to estimate the system state $x(k)$ and the unknown input $u(k)$ from the \textit{only} available system measurement $y(k)$. The main assumption that is imposed to solve this problem is given by Assumption 1 below.

\textit{\underline{Assumptions}}: The system $\mathbf{S}$ is assumed to satisfy the following conditions. namely,   {\textbf{i}) the system is square ($m=l$), \textbf{ii}) the system has a minimal realization, and \textbf{iii}) the system does not have any zeros on the unit circle}.

{In Section \ref{sec:unitcircle} we relax the Assumption (\textbf{iii})}. Other conditions that may be required are provided under each specific statement and result subsequently.

\textit{\underline{\textbf{Problem 2}}: The output tracking:} The objective of this problem is to estimate the input signal $u(k)$ such that the output $y(k)$ follows a desired trajectory $y_d(k)$.  This problem is in fact another re-statement of the Problem 1 above with the difference that the actual output of the system is now replaced by $y_d(k)$. The main assumption that is also required here is Assumption 1.

We now present the notation that is used throughout the paper. Given the matrix $\mathcal{A}$, then $\mathcal{A}^\perp$, $\mathcal{A}^T$ and $\mathcal{N}(\mathcal{A})$ denote the orthogonal space, the transpose, and the null space of $\mathcal{A}$, respectively. We use the concept of \textit{pseudo inverse}.  If $\mathcal{A}$ is full column rank, then we denote the \textit{pseudo inverse} of $\mathcal{A}$ by $\mathcal{A}^\dagger$ and compute it by $(\mathcal{A}^T\mathcal{A})^{-1}\mathcal{A}^T$. If $\mathcal{A}$ is rank deficient, then we denote its pseudo inverse  by $\mathcal{A}^+$, where $\mathcal{A}^+$ is a matrix that satisfies the following four conditions: 1) $\mathcal{A}\mathcal{A}^+\mathcal{A}=\mathcal{A}$, 2) $\mathcal{A}^+\mathcal{A}\mathcal{A}^+=\mathcal{A}^+$, 3) $(\mathcal{A}\mathcal{A}^+)^T=\mathcal{A}\mathcal{A}^+$, and 4) $(\mathcal{A}^+\mathcal{A})^T=\mathcal{A}^+\mathcal{A}$.
If $U\Sigma V^T$ denotes the SVD decomposition of $\mathcal{A}$, then
$\mathcal{A}^+$ is given by $V\Sigma^+U^T$, where $\Sigma^+$ is obtained by
reciprocating each non-zero diagonal element of $\Sigma$.
If $\mathcal{A}$ denotes the system matrix, then $\mathcal{A}^{(1)}$ implies transformation of $\mathcal{A}$ under a standard similarity transformation matrix $\mathbf{T}^{(1)}$. If $x(k)$ denotes a vector, then $\hat{x}(k)$ represent an estimate of $x(k)$. Also, $x^{(1)}(k)$ denotes the transformation of $x(k)$ under the similarity matrix $\mathbf{T}^{(1)}$, i.e. $x^{(1)}(k)=\mathbf{T}^{(1)}x(k)$. Finally, $\textrm{diag}(\mathcal{V})$ denotes a diagonal matrix with elements of the vector $\mathcal{V}$ on its diagonal. Consider the \textit{Rosenbrock System Matrix} defined by,
\begin{equation}
M_R(z)=\left[\begin{array}{cc}zI-A&B\\C&D\end{array}\right]
\end{equation}
if $\textrm{rank}(M_R(z))< n+l$, then $z$ is called a \textit{transmission zero} (or simply the zero) of the system $\mathbf{S}$ or the quadruple $(A,B,C,D)$. The abbreviations MP and NMP stand for minimum phase and non-minimum phase systems, respectively.

\vspace{-4mm}
\section{State and Unknown input reconstruction}\label{sec:state-input-reconstruction}
In this section, we consider and develop methodologies for solving the \underline{Problem 1}.  Let us first set up an unknown input observer (UIO) that generates the state $\eta(k)$ as an \textit{estimate} of  $\mathbf{M}x(k)$ by using only the system measurements $y(k)$, where $\mathbf{M} \in \mathbb{R}^{q \times n}$ is a full row rank matrix to be specified. If $\textrm{rank}(\mathbf{M})=n$, then the system states can be fully reconstructed since $\hat{x}(k)=\mathbf{M}^{-1}\eta(k)$. However, such an $\mathbf{M}$ with rank equal to $n$ does not always exist. In fact,  it turns out that the rank of $\mathbf{M}$ is closely related to the transmission zeros of the system $\mathbf{S}$.

More specifically,  we will show that  $\textrm{rank}(\mathbf{M})=n-\beta$,  where  $\alpha$ and $\beta$  are now representing the number of finite  MP and NMP transmission zeros of the system $\mathbf{S}$, respectively. Clearly, $\alpha+\beta$ is not necessarily equal to $n$.  Our strategy is to first construct an $\mathbf{M}$ having the rank $n-\beta$ by using two to be designed matrices  $\mathbf{M}_0$ and $\mathbf{M}_\#$ that are specified subsequently based on the system $\mathbf{S}$ matrices.  Given $\mathbf{M}$, we then introduce a transformation to partition the system states that can be exactly estimated from those where their estimation is obstructed by the NMP transmission zeros of the system. The estimated states will then serve as inputs to a  causal scheme that estimates the remaining set of the system states.

\vspace{-3mm}
\subsection{Partial or full estimation of the system states}
We start by stating our first formal definition.
\begin{defn}
Assume $\mathbf{M} \in \mathbb{R}^{q \times n}$, where $q \leq n$, is a full row rank matrix. We denote $\eta(k)=\mathbf{M}x(k)$ as a partial or full estimate of the system $\mathbf{S}$ states if $q <n$ or $q=n$, respectively.
\end{defn}
Our goal is to design an unknown input observer (UIO) that estimates $\mathbf{M}x$, where $\mathbf{M} \in \mathbb{R}^{q \times n}$, $q \leq n$, is a full row rank matrix.  We consider the governing dynamics of the unknown input observer (UIO) as follows,
\begin{equation}\label{eq:unknown-observer}
\eta(k-n+1)=\hat{A}\eta(k-n)+F\mathbf{Y}(k-n)
\end{equation}
where,
\begin{equation}
\mathbf{Y}(k-n)=\left[\begin{array}{c}y(k-n+1) \\ y(k-n+2) \\ \vdots \\ y(k)
\end{array}   \right] \in \mathbb{R}^{nl}
\end{equation}
with the matrices {$\hat{A} \in \mathbb{R}^{q \times q}$} and {$F \in \mathbb{R}^{q\times (nl)}$} to be specified subsequently. Our objective is to now  select the matrices $\mathbf{M}$, $\hat{A}$ and $F$ such that $\eta(k)-\mathbf{M}x(k) \rightarrow 0$ as $k \rightarrow \infty$. The output measurement equation of the system $\mathbf{S}$ can be alternatively expressed as,
\begin{equation}\label{eq:y-subspace}
\mathbf{Y}(k-n)=\mathbf{C}_{n}x(k-n)+\mathbf{D}_n\mathbf{U}(k-n)
\end{equation}
where,
\vspace{-3mm}
\begin{equation}\label{eq:cm-dm-def}
\mathbf{C}_n=\left(\begin{array}{c}C\\CA\\ \vdots \\ CA^{n-1}\end{array}\right);\mathbf{D}_{n}=\left(\begin{array}{cccc} D&0& \ldots & 0\\ CB& D& \ldots &0 \\ \vdots & \vdots & \vdots & \vdots \\ CA^{n-2}B&CA^{n-3}B&\ldots & D\end{array}\right)
\end{equation}
{where $\mathbf{C}_n \in \mathbb{R}^{(nl) \times n}$, $\mathbf{D}_n \in \mathbb{R}^{(nl) \times (nm)}$ and $\mathbf{U}(k-n)\in \mathbb{R}^{mn}$} is constructed similar to $\mathbf{Y}(k-n)$ from the input sequence. The state equation of the system $\mathbf{S}$ can be expressed as,
\begin{equation}\label{eq:state-subspace}
x(k-n+1)=Ax(k-n)+B\mathbf{I}_n\mathbf{U}(k-n)
\end{equation}
where $\mathbf{I}_n=\left[\begin{array}{cc} \mathbf{I}_{m \times m} & \mathbf{0}_{m \times (n-m)}\end{array}\right]$. Using the equations (\ref{eq:unknown-observer}), (\ref{eq:y-subspace}) and (\ref{eq:state-subspace}), the unknown input observer error dynamics is now governed by,
\begin{eqnarray}\label{eq:uio-error-dyn}
(\eta-\mathbf{M}x)(k-n+1) &=& \hat{A}(\eta(k-n)-\mathbf{M}x(k-n)) \nonumber \\
&+&(\hat{A}\mathbf{M}-\mathbf{M}A+F\mathbf{C}_n)\mathbf{Y}(k-n) \nonumber \\
&+& (F\mathbf{D}_n-\mathbf{M}B\mathbf{I}_n)\mathbf{U}(k-n)
\end{eqnarray}
It now follows that $\mathbf{M}x$ is accurately estimated if and only if (\textbf{i}) $\hat{A}$ is selected to be a Hurwitz matrix, (\textbf{ii}) $0=\hat{A}\mathbf{M}-\mathbf{M}A+F\mathbf{C}_n$, and (\textbf{iii}) $0=F\mathbf{D}_n-\mathbf{M}B\mathbf{I}_n$.  The conditions (\textbf{i})-(\textbf{iii}) above are the well-known unknown input observer equations that are solvable under certain conditions {\cite{UIO}}. We will show that these conditions have a solution if and only if the system $\mathbf{S}$ is MP. However, this will be obtained under the restrictive requirement that $\mathbf{M}$ should be full rank square matrix. We will show subsequently that a solution for NMP systems  exists if a lower rank matrix $\mathbf{M}$ is considered.

From the condition (\textbf{iii}) it follows that,
\begin{equation}\label{eq:general-F}
F=\mathbf{M}B\mathbf{I}_n\mathbf{D}_n^++\mathbf{K}_n(\mathbf{I}-\mathbf{D}_n\mathbf{D}_n^+)
\end{equation}
where $\mathbf{K}_n \in \mathbb{R}^{nl \times nm}$ is an arbitrary matrix. Let us first  denote by $\hat{A}_0$ and $\mathbf{M}_0$ as solution to $\hat{A}$ and $\mathbf{M}$ that satisfy the conditions (\textbf{i})-(\textbf{iii}) corresponding to $\mathbf{K}_n \equiv 0$. Subsequently, we shall return to the general case where $\mathbf{K}_n$ and $(\mathbf{I}-\mathbf{D}_n\mathbf{D}_n^+)$ are nonzero to obtain another solution to $\mathbf{M}$ that we will denote by $\mathbf{M}_\#$. For now for $\mathbf{M}_0$, we have,
\begin{equation}\label{eq:F-def}
F_0=\mathbf{M}_0B\mathbf{I}_n\mathbf{D}_n^+
\end{equation}
If we substitute $F_0$ from equation (\ref{eq:F-def}) into the condition (\textbf{ii}), we obtain,
\begin{equation}\label{eq:sylvester}
\hat{A}_0\mathbf{M}_0=\mathbf{M}_0(A-B\mathbf{I}_n\mathbf{D}_n^+\mathbf{C}_n)
\end{equation}
Equation (\ref{eq:sylvester}) - which is in fact the Sylvester equation - has $\mathbf{M}_0=0$ as its trivial solution. The non-trivial solution to (\ref{eq:sylvester}) is obtained if  $\mathbf{M}_0$ is considered as the transpose of the left eigenvectors of $\Gamma=(A-B\mathbf{I}_n\mathbf{D}_n^+\mathbf{C}_n)$ and $\hat{A}_0$ as a diagonal matrix of $\Gamma$ eigenvalues. It now follows that the full estimation of the system states by the UIO observer (\ref{eq:unknown-observer}) is obstructed by the NMP transmission zeros of the system due to the fact that the eigenvalues of $(A-B\mathbf{I}_n\mathbf{D}_n^+\mathbf{C}_n)$ contain NMP zeros of  the square system $\mathbf{S}$ as formally stated in the following theorem.
\begin{thm}\label{thm:zeros-of-a-bipkc}
Let Assumption 1 hold, and $\mathcal{V}=\{v_i|i=1,..,p\}$ denote the set of the system $\mathbf{S}$ invariant zeros, and $\mathcal{Z}=\{0, \ldots ,0\}$ that contains $n-p$ zeros. The eigenvalues of $\Gamma=(A-B\mathbf{I}_n\mathbf{D}_{n}^+\mathbf{C}_{n})$ are given by $\mathcal{V} \cup \mathcal{Z}$.
\end{thm}
\vspace{-4mm}
\begin{pf}
Proof is provided in the Appendix \ref{app:zeros-of-a-bipkc}. $\blacksquare$
\end{pf}
\vspace{-4mm}
\begin{rem}
It should be noted that Theorem \ref{thm:zeros-of-a-bipkc} does not hold for non-square systems. The eigenvalues of  $\Gamma$ may or may not coincide with the transmission zeros of $\mathbf{S}$. Each case needs to be then separately investigated, however, once the eigenvalues of  $\Gamma$ are determined, the remaining procedure for obtaining a solution to the conditions (\textbf{i})-(\textbf{iii}) is similar to that of a square system.
\end{rem}
If the system $\mathbf{S}$ has at least one MP transmission zero, or it has less than $n$ NMP zeros (therefore, the set $\mathcal{Z}$ in Theorem \ref{thm:zeros-of-a-bipkc} is not empty), then at least one eigenvalue of  $\Gamma$ is less than 1, which is denoted by $a$. Let us now set $\hat{A}_0=a$. If  $\mathbf{M}_0^T$  is chosen to be the left eigenvector associated with the eigenvalue $a$, then equation (\ref{eq:sylvester}), and consequently conditions (\textbf{i})-(\textbf{iii}) are satisfied even if the system $\mathbf{S}$ has nonzero NMP transmission zeros. In general, we can state the following result.
\begin{lem}\label{lm:sylvester solution}
Let Assumption 1 hold, and $\mathcal{V}=\{v_i|i=1,..,p\}$ denote the set of the system $\mathbf{S}$ invariant zeros, $\mathcal{Z}=\{0, \ldots ,0\}$ that contains $n-p$ zeros, and $\Theta_\alpha$  the set of MP transmission zeros of $\mathbf{S}$. If $\{\Theta_\alpha \cup \mathcal{Z}\} \neq \emptyset$,  then $F_0=\mathbf{M}_0B\mathbf{I}_n\mathbf{D}_n^+$, $\hat{A}_0=\textrm{diag}(\Theta_\alpha \cup \mathcal{Z})$ and $\mathbf{M}_0^T$ that has left eigenvectors of $\Gamma$ associated with $\textrm{diag}(\Theta_\alpha \cup \mathcal{Z})$ are solutions to the conditions (\textbf{i})-(\textbf{iii}).
\end{lem}
\vspace{-4mm}
\begin{pf}
Follows by direct substitution of the solution above into the conditions (\textbf{i})-(\textbf{iii}) that verifies the result. $\blacksquare$
\end{pf}
\vspace{-4mm}
\begin{rem}
One may suggest to use the Jordan canonical form of $\Gamma$ to obtain a solution to the conditions (\textbf{i})-(\textbf{iii}), especially when the system $\mathbf{S}$ has repeated MP transmission zeros. This may yield an $\mathbf{M}$  having higher rank condition as compared to the solution provided by Lemma \ref{lm:sylvester solution} under certain limited cases. However, in general this will not lead to a robust numerical procedure and in most cases the algorithm could fail numerically due to ill-conditioning.
\end{rem}
 Lemma \ref{lm:sylvester solution} implies that a solution for NMP systems exists unless the system $\mathbf{S}$ has exactly $n$ NMP transmission zeros (this is highly unusual in real applications). Our proposed methodology for state estimation problem that will be subsequently discussed requires that $\textrm{rank}(\mathbf{M})=n-\beta$ . However, rank of $\mathbf{M}_0$ that is obtained from Lemma \ref{lm:sylvester solution} is not necessarily equal to $n-\beta$, since $\Gamma$ may have multiple eigenvectors due to repeated eigenvalues and the \textit{generalized eigenvectors} are not a solution to the equation (\ref{eq:sylvester}).

Specifically, the set $\mathcal{Z}$ (as defined in Theorem \ref{thm:zeros-of-a-bipkc}) may have $\alpha_z$ elements sharing the same eigenvectors.  We now consider the term $\mathbf{K}_n(\mathbf{I}-\mathbf{D}_n\mathbf{D}_n^+)\mathbf{C}_n$ in order to obtain linearly independent vectors associated with the elements of $\mathcal{Z}$. If the set $\mathcal{Z}$ is not empty, then it implies that $\mathbf{D}_n$ is rank deficient, and therefore $(\mathbf{I}-\mathbf{D}_n\mathbf{D}_n^+)$ is a nonzero matrix.

Let us now construct $\mathbf{M}_\#$  and $\hat{A}_\#$ such that they satisfy the following Sylvester equation,
\begin{equation}\label{eq:A2-M2}
\hat{A}_\#\mathbf{M}_\#=\mathbf{M}_\#(A-B\mathbf{I}_n\mathbf{D}_n^+\mathbf{C}_n)+\mathbf{K}_n(\mathbf{I}-\mathbf{D}_n\mathbf{D}_n^+)\mathbf{C}_n
\end{equation}
Since $(\mathbf{I}-\mathbf{D}_n\mathbf{D}_n^+)\mathbf{C}_n$ is not identically zero, a non-trivial solution exists and  $\hat{A}_\#$, $\mathbf{M}_\#$ and $\mathbf{K}_n$ can be selected such that the condition (\textbf{i}) is satisfied. Therefore, we have the following theorem.
\begin{thm}\label{thm:complete-solution}
Let Assumption 1 hold and all the MP transmission zeros of $\mathbf{S}$ have an algebraic multiplicity of 1. Then, the complete solution to the conditions (\textbf{i})-(\textbf{iii}) is given by,
\begin{equation}
\hat{A}=\left[\begin{array}{cc}\hat{A}_0 &0 \\ 0 & \hat{A}_\#\end{array}\right];\mathbf{M}=\left[\begin{array}{c} \mathbf{M}_0 \\ \mathbf{M_\#}\end{array}\right]
\end{equation}
where $\textrm{rank}(\mathbf{M})=n-\beta$.
\end{thm}
\vspace{-4mm}
\begin{pf}
\vspace{-4mm}
The proof is provided in the Appendix \ref{app:complete-solution}. $\blacksquare$
\end{pf}
Note that if the system $\mathbf{S}$ has MP transmission zeros with an algebraic multiplicity that is higher than 1, then the rank of $\mathbf{M}_\#$ is reduced proportionally by the multiplicity of the MP transmission zeros. This is  due to the fact that $(\mathbf{I}-\mathbf{D}_n\mathbf{D}_n^+)$ loses its rank. On the other hand, $\mathbf{M}_0$ also loses its rank by such MP transmission zeros. Therefore, our method fails, since the rank of $\mathbf{M}$ will be less than $n-\beta$. {However, we will introduce a technique in Section \ref{sec:unitcircle} to relax the assumption on simplicity of the MP zeros.}

The solution given in equation (\ref{eq:A2-M2}) is closely related to equation (\ref{eq:y-subspace}). The matrix $\mathbf{I}-\mathbf{D}_n\mathbf{D}_n^+$ gives the null space of $\mathbf{D}_n$. Multiplication of both sides of equation (\ref{eq:y-subspace}) by this matrix yields,
\begin{equation}
(\mathbf{I}-\mathbf{D}_n\mathbf{D}_n^+)\mathbf{Y}(k-n)=(\mathbf{I}-\mathbf{D}_n\mathbf{D}_n^+)\mathbf{C}_nx(k-n)
\end{equation}
Let us now define $\mathbf{P}=(\mathbf{I}-\mathbf{D}_n\mathbf{D}_n^+)\mathbf{C}_n$.  It follows that the rank of $\mathbf{P}$ depends on the rank of $\mathcal{N}(\mathbf{D}_n)$. If the system $\mathbf{S}$ has exactly $p=n$ transmission zeros, then $\mathcal{N}(\mathbf{D}_n)=0$, and consequently $\mathbf{P}\equiv 0$. On the other hand, $\mathbf{M}_0$ will be full row rank and will have $n-\beta$ linearly independent rows if the MP transmission zeros are simple. As $p$ is reduced,  then the rank of $\mathbf{P}$ increases and the rank of $\mathbf{M}_0$ decreases. This relationship reveals several important characteristics of $\mathcal{N}(\mathbf{D}_n)$. A more detailed discussion of these properties is beyond the scope of this paper.
\subsection{Partitioning of the states}\label{subsec:partition}
If the system $\mathbf{S}$ has any NMP transmission zeros, then $\textrm{rank}(\mathbf{M})=q<n$, and therefore the states cannot be fully estimated. Let us now perform an LQ decomposition of the matrix $\mathbf{M}$ to partition the estimation of the $q$ states from the estimation of the other $n-q$ states. Namely, let us set $\mathbf{M}=LQ$.

The unknown input observer (UIO) is described by equation (\ref{eq:unknown-observer}), where $\hat{A}$ and $F$ are selected according to Theorem \ref{thm:complete-solution} and equation (\ref{eq:general-F}), and where $\eta(k-n) = \mathbf{M}\hat{x}(k-n)$. Equivalently, we have $\eta(k-n) = LQ\hat{x}(k-n)$. Let us now set the  similarity transformation matrix $\mathbf{T}^{(1)}=Q$. Therefore, $\eta(k-n)=\left[\begin{array}{cc}\mathbf{M}^{(1)}_q&0\end{array}\right]
\hat{x}^{(1)}(k-n)$\footnote{Recall the notation that was defined in Section \ref{sec: problems}, namely, $x^{(1)}(k)=\mathbf{T}^{(1)}x(k)$, $x_1^{(1)}(k)=\mathbf{T}^{(1)}x^{(1)}(k)$, $A^{(1)}=\mathbf{T}^{(1)}A(\mathbf{T}^{(1)})^{-1}$, $B^{(1)}=\mathbf{T}^{(1)}B$, and $C^{(1)}=C(\mathbf{T}^{(1)})^{-1}$.}, where $\left[\begin{array}{cc}\mathbf{M}^{(1)}_q&0\end{array}\right]=L$. The matrix $\mathbf{M}^{(1)}_q \in \mathbb{R}^{q \times q}$ is a non-singular matrix, hence the first $q$ states can be independently reconstructed from $\eta(k-n)$ as follows,
\begin{equation}\label{eq:x1-reconstruction}
\hat{x}^{(1)}(1:q)(k-n)=\mathbf{M}_q^{(1)^{-1}} \eta(k-n)
\end{equation}
where $x(1:q)$ denotes the first $q$ elements of the vector $x$.
\begin{defn}\label{def:mp-nmp}
The MP and NMP states correspond to the first $q$ and the last $n-q$ states of the system $S^{(1)}$ and are denoted by $x_1^{(1)}(k)$ and $x_2^{(1)}(k)$, respectively. In other words, $x^{(1)}(k)=\left[\begin{array}{cc}(x_1^{(1)}(k))^T &(x_2^{(1)}(k))^T
\end{array}\right]^T$, where
\begin{equation}\label{eq:system-s1}
\mathbf{S}^{(1)}: \left\lbrace\begin{array}{l}
x^{(1)}(k+1)=A^{(1)}x^{(1)}(k)+B^{(1)}u(k)\\y(k)=C^{(1)}x^{(1)}(k)+Du(k)
\end{array}\right.
\end{equation}
\end{defn}
\vspace{-4mm}
Considering the Definition \ref{def:mp-nmp} and equation (\ref{eq:x1-reconstruction}), we have,
\begin{equation}\label{eq:x1-reconstruction-2}
\hat{x}^{(1)}_1(k-n)=(\mathbf{M}_q^{(1)})^{-1} \eta(k-n)
\end{equation}
or in the state space representation,
\begin{equation}\label{eq:x1-reconstruction-state}
\left\lbrace\begin{array}{l}
\eta(k-n+1)=\hat{A}\eta(k-n)+F\mathbf{Y}(k-n)\\ \hat{x}^{(1)}_1(k-n)=(\mathbf{M}_q^{(1)})^{-1} \eta(k-n)
\end{array}\right.
\end{equation}
Equation (\ref{eq:x1-reconstruction-state}) shows that the MP states can be independently and accurately estimated from the system measurements. In other words, $\hat{x}_1^{(1)}(k-n)\rightarrow x_1^{(1)}(k-n)$ as $k \rightarrow \infty$. This is due to the fact that according to the error dynamics (\ref{eq:uio-error-dyn}) and conditions (\textbf{i})-(\textbf{iii}), $\eta(k-n)-\mathbf{M}x(k-n) \rightarrow 0$ as $k\rightarrow \infty$. Therefore, $L\hat{x}^{(1)}(k-n)-Lx^{(1)}(k-n) \rightarrow 0$ as $k \rightarrow \infty$, which yields the desired result. An important property of the MP states is now given by the following theorem.
\begin{thm}\label{thm:x1-feature}
Let Assumption 1 hold. Then $x_1^{(1)}(k) \rightarrow 0$ as $k \rightarrow \infty$ if and only if $y(k)=0$ for $k= k_0, k_0+1, \ldots,  \infty$, $k_0>0$.
\end{thm}
\vspace{-4mm}
\begin{pf}
It is known from the state equation of the system (\ref{eq:x1-reconstruction-state}) that $\eta(k)=0$ if and only if $y(k)=0$ ($\Rightarrow \mathbf{Y}(k)=0$) for $k=k_0, k_0+1, \ldots, \infty$, $k_0>0$. On the other hand, $\eta(k)=\mathbf{M}_q^{(1)}\hat{x}^{(1)}_1(k)$. Since $\mathbf{M}_q^{(1)}$ is a nonsingular matrix, it follows that $\hat{x}^{(1)}_1(k)\equiv 0$  if and only if $\eta(k) \equiv 0$. Moreover, $x_1^{(1)}(k) \rightarrow \hat{x}^{(1)}_1(k)$ as $k \rightarrow \infty$. Therefore, $x_1^{(1)}(k) \rightarrow 0$ as $k \rightarrow \infty$, if and only if $y(k)=0$ for $k= k_0, k_0+1, \ldots,  \infty$, $k_0>0$.
$\blacksquare$
\end{pf}
\vspace{-4mm}
The above partitioning is quite helpful in several ways. The most important one is that it renders an elegant expression for the NMP states reconstruction estimation error   as discussed in the next section. Furthermore, in certain applications such as in fault detection and isolation problems, the considered faults may only affect the MP states of the system. Therefore, it will not be necessary to estimate the NMP system states that can be  computationally costly as well as an error prone process.
\vspace{-4mm}
\subsection{Dynamics of the MP and NMP states}
The unknown input estimation problem requires a successful reconstruction of both the MP and the NMP states. Towards this end, we partition the state space model of the system $\mathbf{S}$ or $\mathbf{S}^{(1)}$ as follows ($x_1^{(1)}(k) \in \mathbb{R}^q$ and $x_2^{(1)}(k) \in \mathbb{R}^{n-q}$),
\begin{equation}\label{eq:system-partition}
\mathbf{S}^{(1)}:\left\lbrace\begin{array}{l}x_1^{(1)}(k-n+1)=A_{11}^{(1)}x_1^{(1)}(k-1) \\+A_{12}^{(1)}x_2^{(1)}(k-n)+B_1^{(1)}u(k-n) \\ x_2^{(1)}(k-n+1)=A_{21}^{(1)}x_1^{(1)}(k-n)\\+A_{22}^{(1)}x_2^{(1)}(k-n)+B_2^{(1)}u(k-n)\\ y(k-n)=C_1^{(1)}x_1^{(1)}(k-n)+\\  \left[\begin{array}{cc}C_2^{(1)}&D\end{array}\right]\left[\begin{array}{c}x_2^{(1)}(k-n)\\u(k-n)\end{array}\right]\end{array}\right.
\end{equation}
where,
\begin{equation}\label{eq:A1-partition}
A^{(1)}=\left[\begin{array}{cc} A_{11}^{(1)} & A_{12}^{(1)} \\ A_{21}^{(1)} & A_{22}^{(1)}\end{array}\right];B^{(1)}=\left[\begin{array}{c} B_{1}^{(1)} \\ B_{2}^{(1)}\end{array}\right];C^{(1)}=\left[\begin{array}{cc} C_{1}^{(1)} & C_{2}^{(1)}\end{array}\right].\end{equation}
It is now straightforward to conclude from Theorem \ref{thm:x1-feature} that the following lemmas imply that the NMP states cannot be algebraically estimated from the MP states and the system measurement outputs. Specifically, we have:
\begin{lem}\label{lm:linear-dependence}
Let Assumption 1 hold and  $0<q<n$. Then the columns of $\left[\begin{array}{cc}C_2^{(1)}&D\\ A_{12}^{(1)}&B_1^{(1)}\end{array}\right]$  are linearly dependent.
\end{lem}
\vspace{-4mm}
\begin{pf}
Proof is provided  in the Appendix \ref{app:linear-dependence}.  $\blacksquare$
\end{pf}
\vspace{-4mm}
\begin{lem}\label{lm:nmp-zeros-b}
Let Assumption 1 hold and $0<q<n$. Then the transmission zeros of $\left[\begin{array}{cc}A_{22}^{(1)} & B^{(1)}_2 \\ A_{12}^{(1)} & B_1^{(1)}\end{array}\right]$ are a subset of the system $\mathbf{S}$ transmission zeros.
\end{lem}
\vspace{-4mm}
\begin{pf}
Proof is provided in the Appendix \ref{app:nmp-zeros-b}.  $\blacksquare$
\end{pf}
\vspace{-4mm}
\begin{lem}\label{lm:nmp-zeros-d}
Let Assumption 1 hold and $0<q<n$. Then the transmission zeros of $\left[\begin{array}{cc}A_{22}^{(1)} & B^{(1)}_2 \\ C_2^{(1)} & D\end{array}\right]$ are a subset of the system $\mathbf{S}$ transmission zeros.
\end{lem}
\vspace{-4mm}
\begin{pf}
Proof is provided in the Appendix \ref{app:nmp-zeros-d}.  $\blacksquare$
\end{pf}
\vspace{-4mm}
Let us now assume that $B_{1}^{(1)}$ is full column rank. Then, the unknown input $u(k)$ in terms of the system states is obtained by the first expression of equation (\ref{eq:system-partition}), according to
\vspace{-4mm}
\begin{multline}\label{eq:u}
u(k-n)=B_1^{(1)^\dagger}\left(x_1^{(1)}(k-n+1)-\right.\\ \left. A_{11}^{(1)}x_1^{(1)}(k-n)-A_{12}^{(1)}x_2^{(1)}(k-n)\right)
\end{multline}
By substituting the above equation into the second and third  equations of  (\ref{eq:system-partition})  yields,
\begin{equation}\label{eq:system-sz}
 \left\lbrace \begin{array}{l}
x_2^{(1)}(k-n+1)=A_zx_2^{(1)}(k-n)+B_zX_1^{(1)}(k-n)\\ y(k-n)=C_{z2}x_2^{(1)}(k-n)+C_{z1}X_1^{(1)}(k-n)\end{array}\right.
\end{equation}
where,
\begin{equation}\label{eq:az}
A_z=A_{22}^{(1)}-B_2^{(1)}B_1^{(1)^\dagger}A_{12}^{(1)}
\end{equation}
\begin{equation}\label{eq:bz}
B_z= \left[\begin{array}{cc} B_2^{(1)}B_1^{(1)^\dagger}&A_{21}^{(1)}-B_2^{(1)}B_1^{(1)^\dagger}A_{11}^{(1)}\end{array}\right]
\end{equation}
\begin{equation}\label{eq:nmp-input}
X_1^{(1)}(k-n)=\left[\begin{array}{c} x_1^{(1)}(k-n+1)\\x_1^{(1)}(k-n)\end{array}\right]
\end{equation}
and where $C_{z2}=C^{(1)}_2-DB_1^{(1)^\dagger}A_{12}^{(1)}$ and $C_{z1}=\left[\begin{array}{cc} DB_1^{(1)^\dagger}&C^{(1)}_1-DB_1^{(1)^\dagger}A_{11}^{(1)}\end{array}\right]
$. The quadruple $\Sigma_z:=(A_z,B_z,C_{z1},C_{z2})$ have interesting properties that are related to the transmission zeros of the system $\mathbf{S}$. We are now in a position to state our next result.
\begin{thm}\label{thm:sz-properties}
Let Assumption 1 hold, $0<q<n$ and $B_1^{(1)}$ be a full column rank matrix. Then, the eigenvalues of $A_z$ are the NMP zeros of the system $\mathbf{S}$. Moreover, $C_{z2}=0$.
\end{thm}
\vspace{-4mm}
\begin{pf}
The proof is provided in the Appendix \ref{app:sz-properties}.  $\blacksquare$
\end{pf}
\vspace{-4mm}
\begin{rem}
According to Theorem \ref{thm:sz-properties} and the definition of $C_{z2}$, if $D$ happens to be zero, then, $C_2^{(1)}$ must be zero  which implies  $y(k)=C_1^{(1)}x_1^{(1)}(k)$.  This fact seems to be useful for design of a robust fault detection and isolation scheme,
that is left as a topic of our future research.
\end{rem}
If on the other hand $B_1^{(1)}$ is \textit{not} a full column rank matrix, then let us assume that $D$ is full column rank. In this case, the unknown input in terms of the system states is given by the following expression,
\begin{multline}\label{eq:u-d}
u(k-n)=D^\dagger \left(y(k-n)-C_1^{(1)}x_1^{(1)}(k-n) \right. \\ \left. -C_2^{(1)}x_2^{(1)}(k-n)\right)
\end{multline}
By substituting equation (\ref{eq:u-d}) into the second equation of (\ref{eq:system-partition}), it yields,
\begin{equation}\label{eq:x2-dynamic-d}
x_2^{(1)}(k-n+1)=A_{zd}x_2^{(1)}(k-n)+B_{zd}X_{1d}^{(1)}(k-n)
\end{equation}
where,
\begin{equation}\label{eq:azd}
A_{zd}=A_{22}^{(1)}-B_2^{(1)}D^\dagger C_{2}^{(1)}
\end{equation}
\begin{equation}\label{eq:bzd}
B_{zd}= \left[\begin{array}{cc} A_{21}^{(1)}-B_2^{(1)}D^\dagger C_{1}^{(1)} & B_2^{(1)}D^\dagger\end{array}\right]
\end{equation}
\begin{equation}\label{eq:nmp-input-d}
X_{1d}^{(1)}(k-n)=\left[\begin{array}{c} x_1^{(1)}(k-n)\\y(k-n)\end{array}\right]
\end{equation}
We can now state the next result of this paper.
\begin{thm}\label{thm:sz-properties-d}
Let Assumption 1 hold, $0<q<n$, and $D$ be a full column rank matrix. Then, the eigenvalues of $A_{zd}$ are the NMP zeros of the system $\mathbf{S}$.
\end{thm}
\vspace{-4mm}
\begin{pf}
 Proof is provided in the Appendix \ref{app:sz-properties-d}.  $\blacksquare$
\end{pf}
\vspace{-4mm}
It should be noted that if both $B_1^{(1)}$ and $D$ are column rank deficient matrices, then the NMP states and the unknown input can no longer be estimated. This is a slightly stronger assumption than the \textit{input observability} that requires the matrix $\left[\begin{array}{c}B^{(1)}\\D\end{array}\right]$ to be full column rank.
\vspace{-4mm}
\subsection{Estimation of the NMP states}
The state equation (\ref{eq:system-sz}) (or similarly the equation (\ref{eq:x2-dynamic-d}) depending on the rank condition of $B_1^{(1)}$ or $D$) describes the dynamics of the NMP states. The eigenvalues of $A_z$ (or $A_{zd}$) coincide with the NMP transmission zeros of the system $\mathbf{S}$. Therefore, the dynamics of equation (\ref{eq:system-sz}) or equation (\ref{eq:x2-dynamic-d}) is unstable. This unstable dynamics should be treated in a manner that provides a stable estimation  of the NMP states. Towards this end, let us now consider the following \textit{non-casual} structure that is obtained by re-arranging the state representation (\ref{eq:system-sz}) or (\ref{eq:x2-dynamic-d}) as follows
\begin{equation}\label{eq:noncausal-zero-dynamic-org}
{x}_2^{(1)}(k-n)=\tilde{A}_z{x}_2^{(1)}(k-n+1)-\tilde{B}_z{\Theta}_1^{(1)}(k-n)
\end{equation}
where,
\begin{equation}\label{eq:A-inv}
\tilde{A}_z=(A_z)^{-1} \mbox{ (for (\ref{eq:system-sz})) or } (A_{zd})^{-1} \mbox{ (for (\ref{eq:x2-dynamic-d}))}
\end{equation}
\begin{equation}\label{eq:B-inv}
\tilde{B}_z=(A_z)^{-1}B_z\mbox{ (for (\ref{eq:system-sz})) or }(A_{zd})^{-1}B_{zd} \mbox{ (for (\ref{eq:x2-dynamic-d}))}
\end{equation}
\begin{equation}
\Theta^{(1)}_1(k-n)=X_1^{(1)}(k-n)\mbox{ (for (\ref{eq:system-sz})) or }X_{1d}^{(1)}(k-n)\mbox{ (for (\ref{eq:x2-dynamic-d}))}
\end{equation}
Iterating equation (\ref{eq:noncausal-zero-dynamic-org}) for $n_d$ time steps yields,
\begin{multline}\label{eq:FIR-org}
x_2^{(1)}(k-n-n_d)=\tilde{A}_z^{n_d}x_{2}^{(1)}(k-n)-\\ \sum_{i=0}^{n_d-1} (\tilde{A}_z)^{i}\tilde{B}_z\Theta_1^{(1)}(k-n-i-1)
\end{multline}
where $\tilde{A}_z^{n_d}$ denotes $\tilde{A}_z$ raised to the power of $n_d$. The inverse of $A_z$ (or $A_{zd}$) exists since $A_z$ (or $A_{zd}$) does not have a zero eigenvalue. Also, $\tilde{A}_z$ is Hurwitz due to the fact that the eigenvalues of the inverse matrix is the inverse of the matrix eigenvalues.
Equation (\ref{eq:FIR-org}) provides the key to estimation of the NMP states.

\vspace{-4.5mm}
Let us now construct the following FIR filter,
\begin{multline}\label{eq:nmp-state-reconstruction}
\hat{x}_2^{(1)}(k-n-n_d)=\tilde{A}_z^{n_d}\bar{x}_{20}^{(1)}(k-n)- \\ \sum_{i=0}^{n_d-1} (\tilde{A}_z)^{i}\tilde{B}_z\hat{\Theta}_1^{(1)}(k-n-i-1)
\end{multline}
where $\bar{x}_{20}^{(1)}(k-n)$ denotes the random initial condition of the FIR filter at each time step $k-n$ and $\hat{\Theta}^{(1)}_1(k-n)=\hat{X}_1^{(1)}(k-n)$ or $\hat{\Theta}^{(1)}_1(k-n)=\hat{X}_{1d}^{(1)}(k-n)$, depending on whether  $B^{(1)}$ or $D$ is full column rank, respectively.

Moreover, $\hat{X}_1^{(1)}(k-n)=\left[\begin{array}{c} \hat{x}_1^{(1)}(k-n+1)\\ \hat{x}_1^{(1)}(k-n)\end{array}\right]$ and $\hat{X}_{1d}^{(1)}(k-n)=\left[\begin{array}{c} \hat{x}_1^{(1)}(k-n)\\y(k-n)\end{array}\right]$. The estimate of the MP states ($\hat{x}^{(1)}_1(k)$) as previously discussed is given by (\ref{eq:x1-reconstruction-state}). The random initial condition $\bar{x}_{20}^{(1)}(k-n)$ at each time step introduces errors in the estimation process, but for sufficiently large $n_d$, the effects of the initial conditions will vanish and $\hat{x}_2^{(1)}(k-n-n_d)-x_2^{(1)}(k-n-n_d) \rightarrow 0$ as $k \rightarrow \infty$ (note that  $\tilde{A}_z^{n_d} \rightarrow 0$ for $n_d \gg 1$), as shown subsequently.

Practically, $n_d$ must be as small as possible, however an accurate estimation requires a large $n_d$. Hence, selection of $n_d$ requires a trade-off analysis by quantification of the estimation error versus $n_d$ at each time step. Below, we provide an explicit expression for the reconstruction or estimation error as a function of the delay $n_d$ and the initial condition.
\vspace{-4.5mm}
\begin{defn}\label{def:e-x2}
The NMP state estimation error is defined according to $e_{x2}(k)=x_2^{(1)}(k)-\hat{x}_2^{(1)}(k)$.
\end{defn}
\vspace{-4.5mm}
\begin{thm}\label{thm:reconstruction-error}
Let Assumption 1 hold, $0<q<n$ , and {either $B_1^{(1)}$ or $D$} is a full column rank matrix. Then the NMP state estimation error at the time step $k-n-n_d$ is given by $\tilde{A}_z^{n_d}(x_{2}^{(1)}(k-n)-\bar{x}_{20}^{(1)}(k-n))$.
\end{thm}
\vspace{-4mm}
\begin{pf}
Note that we have,
\begin{eqnarray}
e_{x2}(k-n-n_d)&=&x_2^{(1)}(k-n-n_d)-\hat{x}_2^{(1)}(k-n-n_d) \nonumber \\
&=&\tilde{A}_z^{n_d}(x_{2}^{(1)}(k-n)-\bar{x}_{20}^{(1)}(k-n))\nonumber \\ &-&\sum_{i=0}^{n_d-1} (\tilde{A}_z)^{i}\tilde{B}_z(\Theta_1^{(1)}(k-n-i-1) \nonumber \\
&-&\hat{\Theta}_1^{(1)}(k-n-i-1)) \nonumber
\end{eqnarray}
Since $x_1^{(1)}(k)-\hat{x}_1^{(1)}(k)\rightarrow 0$ as $k \rightarrow \infty$, then $\Theta_1^{(1)}(k)-\hat{\Theta}_1^{(1)}(k) \rightarrow 0$ as $k \rightarrow \infty$. Therefore, the NMP state estimation error is now given by $e_{x2}(k-n-n_d) = \tilde{A}_z^{n_d}(x_{2}^{(1)}(k-n)-\bar{x}_{20}^{(1)}(k-n)) \mbox{ as } k \rightarrow \infty$. $\blacksquare$
\end{pf}
\vspace{-4mm}
Theorem \ref{thm:reconstruction-error} highlights a  number of important trade-off analysis
considerations regarding the nature of the NMP state estimation error and the selection
of the delay $n_d$. Specifically, the following observations can be made:
\begin{itemize}
\item The farther the NMP transmission zeros are from the unit circle,   a smaller NMP state estimation error can be ensured since the term $\tilde{A}_z^{n_d}$ decays faster to zero,
\item The NMP state estimation error for the MP strictly stable system is zero since these systems have a NMP zero at infinity that results in $\tilde{A}_z^{n_d}\equiv 0$, and
\item The closer the NMP transmission zeros are to the unit circle, one can ensure a larger NMP state estimation error to the point that if the system $\mathbf{S}$ has any transmission zeros on the unit circle, then the NMP state estimation results will be certainly biased regardless of the choice of $n_d$.
\end{itemize}
It turns out that one can obtain a conservative upper bound on the NMP state estimation error by considering the 2-norm of $e_{x2}(k)$.  We are now in a position to state our next result.
\begin{thm}\label{thm:error-bound}
Let Assumption 1 hold, $0<q<n$, {either $B_1^{(1)}$ or $D$} is a full column rank matrix and $\bar{x}_{20}^{(1)}(k)=0$ for all $k$. Then $\sup (\|e_{x2}(k)\|_2)=\sigma_{max}(\tilde{A}_z^{n_d})\|(z\mathbf{I}-A^{(1)})^{-1}B^{(1)}\|_\infty$, where $\sigma_{max}(.)$ denotes the largest singular value operator.
\end{thm}
\vspace{-4mm}
\begin{pf}
It follows from Theorem \ref{thm:reconstruction-error} that,
\begin{eqnarray}
\|e_{x2}(k-n-n_d)\|_2&=&\|\tilde{A}_z^{n_d}x_{2}^{(1)}(k-n)\|_2  \\
&\leq & \|\tilde{A}_z^{n_d}\|_2\|x_{2}^{(1)}(k-n)\|_2 \nonumber  \nonumber  \\
& \leq& \sigma_{max}(\tilde{A}_z^{n_d})\|(z\mathbf{I}-A^{(1)})^{-1}B^{(1)}\|_\infty \nonumber
\end{eqnarray}
The last inequality holds since the $L_2$ input-output gain is bounded by the $\infty$-norm of the system $\mathbf{S}$.  $\blacksquare$
\end{pf}
\vspace{-4mm}
The above upper bound can be plotted as a function of $n_d$ to perform a trade-off analysis. Note that $\sigma_{max}(\tilde{A}_z^{n_d})$ is determined by the smallest NMP transmission zero of the system $\mathbf{S}$ due to the fact that the eigenvalues of $\tilde{A}_z$ are inverse of the system $\mathbf{S}$ NMP transmission zeros. This is in accordance with the results that are stated in \cite{Marro2010815}. Note that $\sigma_{max}(\tilde{A}_z^{n_d})$ asymptotically decays  to zero as $n_d$ is increased. Therefore, an almost perfect estimation can be achieved when $n_d$  is equal to several times that of the system order.
\begin{rem}\label{rm:small-nd}
If the system $\mathbf{S}$ is stimulated by an input such that $u(k+1)\neq u(k)$ at finite $k$'s (such as in a step function) or $\|u(k+1)-u(k)\|$ is sufficiently small  (such as in a harmonic function), then one can choose $\bar{x}_{20}^{(1)}(k-n)=\hat{x}_2^{(1)}(k-n-n_d-1)$ in the filter (\ref{eq:nmp-state-reconstruction}) which \underline{may} yield an almost unbiased state estimate by selecting the smallest possible choice of $n_d=2$. This is due to the fact that in these cases $\hat{x}_2^{(1)}(k-n-n_d-1)$ is a close approximation to $x_2^{(1)}(k-n-n_d-1)$ and $x_2^{(1)}(k-n)$ (for small $n_d$), and therefore it may yield a sufficiently small NMP state estimation error, i.e., $e_{x2}(k-n-n_d)=\tilde{A}_z^{n_d}(x_{2}^{(1)}(k-n)-\bar{x}_{20}^{(1)}(k-n))=\tilde{A}_z^{n_d}(x_{2}^{(1)}(k-n)-\hat{x}_2^{(1)}(k-n-n_d-1)) \approx 0$ even if $n_d$ is selected to be sufficiently small.
\end{rem}
We will illustrate the above statement in our simulation case study section. Once both the MP and NMP states are estimated,  the unknown input can now be easily estimated by using equation (\ref{eq:u}) (or (\ref{eq:u-d})). Specifically, if $B^{(1)}_1$ is full column rank, then $\hat{u}(k)$ is given by,
\begin{multline}\label{eq:uhat-b1}
\hat{u}(k-n)=B_1^{(1)^\dagger}\left(\hat{x}_1^{(1)}(k-n+1)- \right. \\ \left. A_{11}^{(1)}\hat{x}_1^{(1)}(k-n)-A_{12}^{(1)}\hat{x}_2^{(1)}(k-n)\right)
\end{multline}
and if $D$ is full column rank, it is given by,
\begin{multline}\label{eq:uhat-d}
\hat{u}(k-n)=D^\dagger \left(y(k-n)-C_1^{(1)}\hat{x}_1^{(1)}(k-n) \right. \\ \left. -C_2^{(1)}\hat{x}_2^{(1)}(k-n)\right)
\end{multline}
\vspace{-4mm}
\begin{defn}\label{def:input-est-error}
The unknown input estimation error is defined according to $e_u(k)=\hat{u}(k)-u(k)$.
\end{defn}
\vspace{-4mm}
\begin{prop}\label{prop:input-reconstruction error-B1}
Let Assumption 1 hold, $0<q<n$, and $B_1^{(1)}$ be a full column rank matrix. Then,
\begin{equation}
e_u(k)\rightarrow-B_1^{(1)^\dagger}A_{12}^{(1)}e_{x2}(k)\mbox{ as } k\rightarrow \infty.
\end{equation}
\end{prop}
\vspace{-6mm}
\begin{pf}
The result follows readily from equations (\ref{eq:u}) and (\ref{eq:uhat-b1}) by noting that $u(k)-\hat{u}(k)=B_1^{(1)^\dagger}(x_1^{(1)}(k+1)-A_{11}^{(1)}x_1^{(1)}(k)-A_{12}^{(1)}x_2^{(1)}(k))-B_1^{(1)^\dagger}(\hat{x}_1^{(1)}(k+1)-A_{11}^{(1)}\hat{x}_1^{(1)}(k)-A_{12}^{(1)}\hat{x}_2^{(1)}(k))\rightarrow-B_1^{(1)^\dagger}A_{12}^{(1)}e_{x2}(k)$ as $k \rightarrow \infty$. This follows due to the fact that $\hat{x}_1^{(1)}(k) \rightarrow x_1^{(1)}(k)$ as $k \rightarrow \infty$ and $e_{x2}(k)=x_2^{(1)}(k)-\hat{x}_2^{(1)}(k)$ (Definition \ref{def:e-x2}).  $\blacksquare$
\end{pf}
\vspace{-4mm}
The Proposition \ref{prop:input-reconstruction error-B1} links  the unknown input estimation error to the state estimation error. This may serve as a means for conducting a trade-off analysis. The above implies that the state estimation error is propagated through the gain $-B_1^{(1)^\dagger}A_{12}^{(1)}$ to the unknown input estimation error. One can interestingly conclude that if $-B_1^{(1)^\dagger}A_{12}^{(1)}$ happens to be zero, then the unknown input estimation process will be unbiased regardless of the NMP states estimation error. Therefore, it can immediately be concluded that $-B_1^{(1)^\dagger}A_{12}^{(1)}=0$  if and only if the NMP zero of the system $\mathbf{S}$ is at infinity. In other words, the system $\mathbf{S}$ is strictly stable and MP.  The proposition \ref{prop:input-reconstruction error-B1}  provides an explicit unknown input estimation error expression if  $B_1^{(1)}$ is full column rank. In case that $D$ is a full column rank matrix, we arrive at the !
 following result.
\vspace{-4mm}
\begin{prop}\label{prop:input-reconstruction error-D}
Let Assumption 1 hold, $0<q<n$ , and $D$ be a full column rank matrix. Then,
\begin{equation}
e_u(k)\rightarrow-D^\dagger C_2^{(1)}e_{x2}(k) \mbox{ as }k \rightarrow \infty.
\end{equation}
\end{prop}
\vspace{-4mm}
\begin{pf}
It follows readily from equations (\ref{eq:u-d}) and (\ref{eq:uhat-d}) that we have $u(k)-\hat{u}(k)=D^\dagger (y(k)-C_1^{(1)}x_1^{(1)}(k)-C_2^{(1)}x_2^{(1)(k)})-D^\dagger (y(k)-C_1^{(1)}\hat{x}_1^{(1)}(k)-C_2^{(1)}\hat{x}_2^{(1)}(k))\rightarrow-D^\dagger C_2^{(1)}e_{x2}(k)$ as $k\rightarrow \infty$. This follows due to the fact that $\hat{x}_1^{(1)}(k) \rightarrow x_1^{(1)}(k)$ as $k \rightarrow \infty$ and $e_{x2}(k)=x_2^{(1)}(k)-\hat{x}_2^{(1)}(k)$ (Definition \ref{def:e-x2}).  $\blacksquare$
\end{pf}
\vspace{-4mm}
An immediate conclusion from the Propositions \ref{prop:input-reconstruction error-B1} and \ref{prop:input-reconstruction error-D} is that if the system $\mathbf{S}$ is NMP and both $B_1^{(1)}$ and $D$ are full column rank matrices, then $B_1^{(1)^\dagger}A_{12}^{(1)}=D^\dagger C_2^{(1)}$,  which we have already derived through a different method in Theorem \ref{thm:sz-properties} ($C_{z2}^{(1)}=0$).
This completes our solution to the \underline{Problem 1}. In the next section, we discuss a solution to the \underline{Problem 2}.

{It is worth pointing that our proposed methodology is not suitable for systems when all the zeros are NMP \underline{and} the system has the same number of poles
and zeros. In fact, this scenario for a square system implies that
the matrix $D$ is full rank, therefore, other methods such as the one in \cite{wahls} are available to handle this particular case.}

\vspace{-4mm}
\section{Inversion-based output tracking}\label{sec:output-tracking}
We have shown earlier that in  presence of NMP states, accurate estimation of the MP states as well as bounded error estimation of the NMP states are possible under certain conditions. In this section, by utilizing the previous results we will introduce and develop an inversion-based output tracking control methodology as a solution to \underline{Problem 2}. Specifically, we will obtain relationship between the resulting tracking error performance and the unknown input and state estimation errors. We also demonstrate that almost perfect tracking of an arbitrary desired output trajectory can be achieved.

For the output tracking problem a delayed state and input estimation may not be useful or practical given that the controller should issue the command at a given present time. This challenge can be resolved if we assume that the desired output trajectory  from $y_d(k)$ to $y_d(k+n+n_d)$ is known \textit{ a priori} at a given time step $k$,  which is known as the \textit{preview time} (window) in the literature \cite{zou1999preview}. This is actually a reasonable and acceptable assumption given that the desired trajectory is typically planned in advance and at least it can be assumed practically to be known for  $n+n_d$ time steps ahead. Our proposed estimation scheme is now slightly modified to incorporate this conditional change. A summary of the procedure for implementation of our proposed scheme is presented in Table \ref{tab:output-tracking-algorithm}.

Let us now define $\mathbf{Y}_d(k)$ as $\left[\begin{array}{ccc}y_d(k)^T & \ldots &  y_d(k+n)^T\end{array}\right]^T$, where $\mathbf{Y}_d(k)$ is assumed to be a \textit{known} signal. It is now utilized to derive the unknown input observer following equation (\ref{eq:x1-reconstruction-state}) to yield $\hat{x}_1^{(1)}(k)$ as follows,
\begin{equation}\label{eq:outputTracking-x1}
 \left\lbrace\begin{array}{l}
\eta(k+1)=\hat{A}\eta(k)+F\mathbf{Y}_d(k)\\  \hat{x}^{(1)}_1(k)=\mathbf{M}_q^{(1)^{-1}} \eta(k)
\end{array}\right.
\end{equation}
An estimate of ${x}_2^{(1)}(k)$ is now given by,
\begin{equation}\label{eq:outputTracking-x2}
\hat{x}_2^{(1)}(k)=\tilde{A}_z^{n_d}\bar{x}_{20}^{(1)}(k+n_d)-\sum_{i=0}^{n_d-1} (\tilde{A}_z)^{i}\tilde{B}_z\hat{\Theta}_1^{(1)}(k+n_d-i-1)
\end{equation}
where $\bar{x}_{20}^{(1)}(k+n_d)$ is a random initial condition of the FIR filter at each time step $k+n_d$, and $\hat{\Theta}^{(1)}_1(k)=\hat{X}_1^{(1)}(k)$ or $\hat{\Theta}^{(1)}_1(k)=\hat{X}_{1d}^{(1)}(k)$, if $B^{(1)}$ or $D$ is full column rank, respectively. Moreover, $\hat{X}_1^{(1)}(k)=\left[\begin{array}{c} \hat{x}_1^{(1)}(k+1)\\ \hat{x}_1^{(1)}(k)\end{array}\right]$ and $\hat{X}_{1d}^{(1)}(k)=\left[\begin{array}{c} \hat{x}_1^{(1)}(k)\\y(k)\end{array}\right]$. If $B^{(1)}_1$ is full column rank, then $\hat{u}(k)$ is given by,
\begin{equation}\label{eq:outputTracking-uhat-b1}
\hat{u}(k)=B_1^{(1)^\dagger}\left(\hat{x}_1^{(1)}(k+1)-A_{11}^{(1)}\hat{x}_1^{(1)}(k)-A_{12}^{(1)}\hat{x}_2^{(1)}(k)\right)
\end{equation}
and if $D$ is full column rank, then $\hat{u}(k)$ is given by,
\begin{equation}\label{eq:outputTracking-uhat-d}
\hat{u}(k)=D^\dagger \left(y(k)-C_1^{(1)}\hat{x}_1^{(1)}(k)-C_2^{(1)}\hat{x}_2^{(1)}(k)\right)
\end{equation}
Since the NMP state estimation scheme is subject to errors, if the computed $\hat{u}(k)$ is fed to the system, it will then generate ${y}(k)$ that is different from the desired $y_d(k)$. In other words, ${y}(k)$ is the real output of the system subjected to and stimulated by $\hat{u}(k)$, that is (in view of the state space representation (\ref{eq:system-s1}))
\begin{equation}\label{eq:system-tilde}
\left\lbrace\begin{array}{l}
\tilde{{x}}^{(1)}(k+1)=A^{(1)}\tilde{{x}}^{(1)}(k)+B^{(1)}\hat{u}(k)\\ {y}(k)=C^{(1)}\tilde{{x}}^{(1)}(k)+D\hat{u}(k)
\end{array}\right.
\end{equation}
where $\tilde{x}^{(1)}(k)$ denotes the state response of the system to the input $\hat{u}(k)$. If the exact $u(k)$ is known, then we would have obtained,
\begin{equation}\label{eq:system-real}
\left\lbrace\begin{array}{l}
{{x}}^{(1)}(k+1)=A^{(1)}{{x}}^{(1)}(k)+B^{(1)}{u}(k)\\ {y}_d(k)=C^{(1)}{{x}}^{(1)}(k)+D{u}(k)
\end{array}\right.
\end{equation}
We are now in a position to define the output tracking error as follows.

\vspace{-2mm}
\begin{defn}\label{def:outputTracking-error}
The output tracking error is defined as $e_y(k)=y(k)-{y}_d(k)$.
\end{defn}
\vspace{-4mm}
It now follows from equations (\ref{eq:system-tilde}) and (\ref{eq:system-real}) that,
\begin{equation}\label{eq:ey-vs-eu}
\left\lbrace\begin{array}{l}
\tilde{{e}}_x(k+1)=A^{(1)}\tilde{{e}}_x(k)+B^{(1)}e_u(k)\\ e_y(k)=C^{(1)}\tilde{{e}}_x^{(1)}(k)+De_u(k)
\end{array}\right.
\end{equation}
where $\tilde{e}_x(k)=\tilde{x}^{(1)}(k)-{x}^{(1)}(k)$. It is straightforward to conclude from equation (\ref{eq:ey-vs-eu}) that $e_y(k) \rightarrow 0$ as $k \rightarrow \infty$ if $e_u(k) \rightarrow0$ as $k \rightarrow \infty$. However, $e_u(k)$ is given by the NMP state estimation error ($e_{x2}(k)$) that is multiplied by a gain as formally stated in Propositions \ref{prop:input-reconstruction error-B1} or \ref{prop:input-reconstruction error-D}. We have shown in Theorem \ref{thm:error-bound} that the NMP state estimation error ($e_{x2}(k)$) decays asymptotically as $n_d$ increases. Hence, an almost perfect output tracking for any desired trajectory can be achieved by selecting $n_d$ to be sufficiently large by as much as few times of the system order in most cases. The following theorem formally establishes the above statement and provide an upper bound on the output tracking error versus the delay parameter $n_d$.
\vspace{-4mm}
\begin{thm}\label{thm:output-tracking-error}
Let Assumption 1 hold and $0<q<n$. If $B^{(1)}_1$ is full column rank, then $\|e_y(k)\|_2 \leq \sigma_{max}(\tilde{A}_z^{n_d})\|C(zI-A)^{-1}B+D\|_\infty\|B_1^{(1)^\dagger}A_{12}^{(1)}\|_2\|(z\mathbf{I}-A^{(1)})^{-1}B^{(1)}\|_\infty$. On the other hand, if $D$ is full column rank, then $\|e_y(k)\|_2 \leq \sigma_{max}(\tilde{A}_z^{n_d})\|C(zI-A)^{-1}B+D\|_\infty\|D^{\dagger}
C_{2}^{(1)}\|_2\|(z\mathbf{I}-A^{(1)})^{-1}B^{(1)}\|_\infty$.
\end{thm}
\vspace{-4mm}
\begin{pf}
According to equation (\ref{eq:ey-vs-eu}), $\|e_y(k)\|_2=\|C(zI-A)^{-1}B+D\|_\infty\|e_u(k)\|_2$. If $B^{(1)}_1$ is full column rank, then from Proposition \ref{prop:input-reconstruction error-B1}, $\|e_u(k)\|_2=\|B_1^{(1)^\dagger}A_{12}^{(1)}e_{x2}(k)\|_2 \leq \|B_1^{(1)^\dagger}A_{12}^{(1)}\|_2\|e_{x2}(k)\|_2$. Our desired result is now obtained if we substitute  $\|e_{x2}(k)\|_2$, by using Theorem \ref{thm:error-bound}, into the above expression  as $\sigma_{max}(\tilde{A}_z^{n_d})\|(z\mathbf{I}-A^{(1)})^{-1}B^{(1)}\|_\infty$. Following along the same procedure yields our other desired result for the case when $D$ is full column rank.  $\blacksquare$
\end{pf}
\vspace{-4mm}
As expected, Theorem \ref{thm:output-tracking-error} implies that the upper bound of the output tracking error has the same functionality in terms of the delay parameter $n_d$ as that of the upper bound of the NMP states estimation error. Theorem \ref{thm:output-tracking-error} is quite useful for performing a trade-off analysis between the delay parameter $n_d$ and the output tracking error. This completes our proposed methodology for inversion-based output tracking. In the next section, we show that our proposed solution provides a framework for systematic treatment of the systems with transmission zeros on the unit circle.
\begin{table}
\centering
\caption{Inversion-based output tracking algorithm.}
\label{tab:output-tracking-algorithm}
\begin{tabular}{|p{0.45\textwidth}|}
 \hline
\begin{enumerate}
\item Calculate $\hat{A}$, $F$ and $\mathbf{M}$ from Theorem \ref{thm:complete-solution}.
\item Calculate $\mathbf{T}^{(1)}$ and $\mathbf{M}_q^{(1)}$ from equations given in Section 3.2.
\item Calculate $A^{(1)}$, $B^{(1)}$ and $C^{(1)}$ by applying the similarity transformation to the system $\mathbf{S}$ using the matrix $\mathbf{T}^{(1)}$ ($x^{(1)}=\mathbf{T}^{(1)}x$).
\item Partition $A^{(1)}$, $B^{(1)}$ and $C^{(1)}$ according to equation (\ref{eq:A1-partition}).
\item If $B^{(1)}_1$ is full column rank, then obtain $A_z$ and $B_z$ from equations (\ref{eq:az}) and (\ref{eq:bz}). If $B^{(1)}_1$ is not full column rank and $D$ is full column rank, then obtain $A_{zd}$ and $B_{zd}$ from equations (\ref{eq:azd}) and (\ref{eq:bzd}).
\item Calculate  $\tilde{A}_z$ and $\tilde{B}_z$ from equations (\ref{eq:A-inv}) and (\ref{eq:B-inv}).
\item Select $n_d$  according to Theorem \ref{thm:output-tracking-error} to meet the desired estimation error specifications .
\item At each time step $k$,
\begin{enumerate}
\item Reconstruct $\hat{x}_1^{(1)}(k)$ from equation (\ref{eq:outputTracking-x1}).
\item Reconstruct $\hat{x}_2^{(1)}(k)$ using equation (\ref{eq:outputTracking-x2}).
\item If $B^{(1)}_1$ is full column rank, then reconstruct $\hat{u}(k)$ from equation (\ref{eq:outputTracking-uhat-b1}). If $B^{(1)}_1$ is not full column rank and $D$ is full column rank, then reconstruct $\hat{u}(k)$ from equation (\ref{eq:outputTracking-uhat-d}).
\end{enumerate}
\end{enumerate}
 \\ \hline
\end{tabular}
\end{table}
{
\section{Systems having transmission zeros on the unit circle}\label{sec:unitcircle}
In this section, we show that our proposed approach can be extended and applied for handling the output tracking problem in systems that have transmission zeros on the unit circle in addition to MP and NMP zeros. This problem has \underline{not} been addressed and solved in the literature. For simplicity of the discussion, we only consider a SISO system that is described by,
\begin{eqnarray}\label{eq:gz-def}
Y(z)=G(z)U(z) = (z+1)G'(z)U(z)
\end{eqnarray}
The function $G(z)$ is given by the z-transform of the system $\mathbf{S}$ that has a transmission zero at $z=-1$, and therefore can be written as $(z+1)G'(z)$, where all the zeros and poles of $G'(z)$ are inside the unit circle. The extension of our proposed solution to a general MIMO system with non-minimum phase transmission zeros and multiple zeros on the unit circle is straightforward and not included here for simplicity. Our objective  is to determine $u(k)$ such that $y(k)$ follows the desired and known trajectory $y_d(k)$. Note that if such a $u(k)$ is found, it will also satisfy $Y_d(z)=G(z)U(z)$, where $Y_d(z)$ denotes the z-transform of $y_d(k)$. \\}

{Let us rewrite $Y_d(z)$ as $Y_d(z)=zY'_d(z)+Y'_d(z)$, where $Y'_d(z)=G'(z)U(z)$. Clearly, $Y'_d(z)$ is given by
\begin{equation}\label{eq:ydz}
Y'_d(z)=(z+1)^{-1}Y_d(z)
\end{equation}
The exact computation of $y'_d(k)=\mathcal{Z}^{-1}\{Y'_d(z)\}$ requires the initial condition $y'_d(0)$, which is not known. If we rewrite equation (\ref{eq:ydz}) in the time-domain, we have,
\begin{equation}\label{eq:ydz2}
y'_d(k+1)=-y'_d(k)+y_d(k)
\end{equation}
Therefore, we have $y'_d(1)=-y'_d(0)+y_d(0)$ at $k=0$. Although $y_d(0)$ is known, but $y'_d(0)$ is unknown. On the other hand, the pole of the system (\ref{eq:ydz2}) is at -1, thus the effect of unknown initial condition will not die out overtime. If $y'_d(0)$ was known, since $U(z)$ satisfies both $Y_d(z)=G(z)U(z)$ and $Y'_d(z)=G'(z)U(z)$, then one would simply use the algorithm of Table \ref{tab:output-tracking-algorithm} to compute $u(k)$ by using $G'(z)$ and the exact value of $y'_d(k)$ instead of $G(z)$ and $y_d(k)$. Consequently, the problem of handling transmission zeros on the unit circle would be easily resolved. In fact, the Algorithm provided in Table \ref{tab:output-tracking-algorithm}, similar to the other work, e.g. in \cite{kirtikar2009delay} can generate an almost exact duplicate of the unknown states and inputs that are biased by the value of the unknown initial condition. However, our objective here is to diminish the effects of unknown initial conditions !
 so that the system outputs converge to the desired trajectory starting from an arbitrary initial condition. }

{Let us assume that the system $\mathbf{S}'$ is specified  by the quadruple $\Sigma':=(A',B',C',D')$ and goverend by $\mathcal{Z}^{-1}(G'(z))$. Note that the system $\mathbf{S}'$ is minimum phase. Therefore, according to the equation (\ref{eq:x1-reconstruction-state}), the estimate of all the system states is given by,
\begin{equation}\label{eq:xz}
\hat{X}(z)=(z\mathbf{I}-\mathcal{A})^{-1}\mathcal{F}\mathtt{Z}_MY'_d(z)
\end{equation}
where $\mathcal{A}=\left(\mathbf{M}_q^{(1)}\right)^{-1}\hat{A}\mathbf{M}_q^{(1)}$, $\mathcal{F}=\left(\mathbf{M}_q^{(1)}\right)^{-1}F$ and $\mathtt{Z}_M=\left[\begin{array}{ccc}z^{0}&\ldots&z^{n-1}\end{array}\right]^T$, where $z^{-1}$ denotes the delay operator. The matrices $\mathbf{M}_q^{(1)}$, $\hat{A}$ and $F$ are computed from Theorem \ref{thm:complete-solution} by using the quadruple $\Sigma'$. The transfer function in equation (\ref{eq:xz}) is stable since the poles of $\mathcal{A}$ are the MP zeros of the system. }

\vspace{-3mm}
{Let us introduce the controller $H(z)$ having the order $n_c$ that can be written as $(z+1)H'(z)$. It is necessary for $H(z)$ to have $(z+1)$ as a factor in order to cancel out the pole of the inverse system at $z=-1$. Moreover, let us assume that an approximation to the system $\mathbf{S}'$ states is given by $\tilde{X}(z)=H(z)\hat{X}(z)$. Therefore, according to equation (\ref{eq:uhat-b1}), an approximation $\tilde{U}(z)$ to the  unknown input is given by,
\begin{equation}\label{eq:uz}
\tilde{U}(z)=(B'^{(1)})^\dagger (z\mathbf{I}-A'^{(1)})H(z)\hat{X}(z)
\end{equation}
where $B'^{(1)}=QB'$ and $A'^{(1)}=Q{A'}Q^{-1}$. Combining equations (\ref{eq:ydz}), (\ref{eq:xz}) and (\ref{eq:uz}) yield,
\begin{equation}\label{eq:uz-h}
\tilde{U}(z)=(B'^{(1)})^\dagger (z\mathbf{I}-A'^{(1)})H(z)(z\mathbf{I}-\mathcal{A})^{-1}\mathcal{F}\mathtt{Z}_M(z+1)^{-1}Y_d(z)
\end{equation}
By canceling out the pole at $z=-1$ by the $(z+1)$ factor of $H(z)$, one obtains,
\begin{eqnarray}\label{eq:uz-hp-bias}
\tilde{U}(z)&=&(B'^{(1)})^\dagger (z\mathbf{I}-A'^{(1)})H'(z)(z\mathbf{I}-\mathcal{A})^{-1}\mathcal{F}\mathtt{Z}_MY_d(z) \nonumber \\
&=&\tilde{G}_i(z)Y_d(z)
\end{eqnarray}
Equation (\ref{eq:uz-hp-bias}) provides a direct and biased estimate of the system input by using $y_d(k)$ with an additional delay of $n_c$. Theoretically, the unbiased estimate of the system input is given by ($H(z)=\mathbf{I}$),
\begin{eqnarray}\label{eq:uz-hp}
\hat{U}(z)&=&(B'^{(1)})^\dagger (z\mathbf{I}-A'^{(1)})(z\mathbf{I}-\mathcal{A})^{-1}\mathcal{F}\mathtt{Z}_M(z+1)^{-1}Y_d(z) \nonumber \\
&=&\hat{G}_i(z)Y_d(z)
\end{eqnarray}
Therefore, the output tracking error is approximated by,
\begin{eqnarray}\label{eq:error-uhz}
E_c(z)&=&\hat{Y}(z)-\tilde{Y}(z) \nonumber \\
&=&G(z)(\hat{G}_i(z)-\tilde{G}_i(z))Y_d(z)
\end{eqnarray}
The above result provides an important criterion for selecting the controller $H(z)$. Note that the governing error dynamics (\ref{eq:error-uhz}) does not have a pole at $z=-1$ since the pole of $\hat{G}_i(z)$ has been canceled out by the $(z+1)$ factor of $G(z)$. Therefore, the system (\ref{eq:error-uhz}) is stable. \\}

\vspace{-3mm}
{Let us assume that for the particular case we considered here the function $H(z)$ is expressed as $\mathcal{H}(z)\mathbf{I}$, where $\mathcal{H}(z)$ represents a SISO transfer function of a controller that contains a factor $(z+1)$. Therefore, the error dynamics (\ref{eq:error-uhz}) is given by,
\begin{equation}
E_c(z)=G(z)\mathcal{G}_i(z)(1-\mathcal{H}(z))Y_d(z)
\end{equation}
where $\mathcal{G}_i(z)=(B'^{(1)})^\dagger (z\mathbf{I}-A'^{(1)})(z\mathbf{I}-\mathcal{A})^{-1}\mathcal{F}\mathtt{Z}_M$. Therefore, the design problem is reduced to that of solving the following optimization problem,
\begin{equation}
\min_{\mathcal{H}(z)}\|1-\mathcal{H}(z)\|
\end{equation}
which is also known as the $H_\infty$ (or $H_2$) norm minimization problem. }

{Clearly, the above problem is associated with numerous trade-off considerations. For instance, one may not be able to achieve a minimum value over all frequencies. One may directly introduce the controller $\mathcal{H}(z)$ by defining an approximation of the system input given by $\tilde{U}(z)=\mathcal{H}(z)\hat{U}(z)$. The choice depends on the design preference requirements. Using $H(z)$ provides  more degrees of freedom, however it also complicates the minimization problem and the trade-off studies.}

{A significant advantage of our solution is derived from the fact that both minimum and non-minimum phase zeros will not be involved in the design process of $H(z)$ to handle the transmission zeros on the unit circle. In other words, the controller $H(z)$ should only cancel
out the transmission zeros on the unit circle, and therefore one can always
find such a controller. However, the actual challenge is due to the fact that the controller
should also minimize the $\|1-{H}(z)\|$ over a desired range of
frequencies.}

{Let us further assume that the particular case we consider here  has also a non-minimum phase transmission zero.  In this case, we exactly follow the algorithm that is provided in Table \ref{tab:output-tracking-algorithm} subject to two modifications, namely: \textbf{i}) we use the quadruple $\Sigma'$ to determine the algorithm parameters, and \textbf{ii}) we use the following relationship for estimation of the MP states instead of equation (\ref{eq:x1-reconstruction-state}), namely
\[\tilde{X}(z)=H'(z)(z\mathbf{I}-\mathcal{A})^{-1}\mathcal{F}\mathtt{Z}_MY_d(z)\]
Therefore, the inverse dynamics poles are the same as the poles of $H'(z)$ and  $\mathcal{A}$ which are inside the unit circle. Note that the order of $\mathcal{A}$ is less than $n$, therefore $H'(z)$ should be selected such that the above transfer function becomes proper. }

{Finally, following along a similar approach allows one to relax the condition of simplicity of the MP transmission zeros in Theorem \ref{thm:complete-solution}. As an illustration, let us assume that the transfer function of a SISO system can be expressed as $G(z)=(z-p)^2G''(z)$, where $\|p\|<1$ and $G''(z)$ does not have $(z-p)$ as a factor in its denominator.  Therefore, the algebraic multiplicity of $z=p$ is two. In this case, we exactly follow the algorithm that is provided in Table \ref{tab:output-tracking-algorithm} subject to two modifications, namely: \textbf{i}) we use the quadruple $\Sigma'$ as described by $\mathcal{Z}^{-1}\{(z-p)G''(z)\}$ to determine the algorithm parameters, and \textbf{ii}) we use the signal $y'_d(k)$ instead of $y_d(k)$ where $y'_d(k)$ is governed by $Y'_d(z)=\frac{Y_d(z)}{z-p}$. The above transfer function has an arbitrary initial condition. Since $\|p\|<1$, the effects of the unknown initial condition will die out quickly. !
 The solution also does not require the introduction of a controller.\\}

{This now completes our proposed methodology for handling transmission zeros on the unit circle. In the next section, we will provide illustrative simulations to demonstrate the merits of our proposed methodologies.}
\vspace{-4mm}
\section{Numerical Case Studies Simulations}\label{sec:simulation}
\begin{figure}
  \centering
  \includegraphics[width=0.47\textwidth]{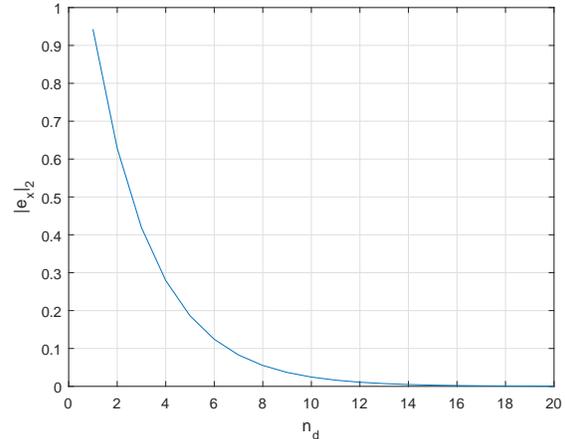}\\
\vspace{-4mm}
  \caption{Upper bound of the NMP state estimation error versus $n_d$.}\label{fig:nd}
\end{figure}

\begin{figure}
  \centering
  \includegraphics[width=0.47\textwidth]{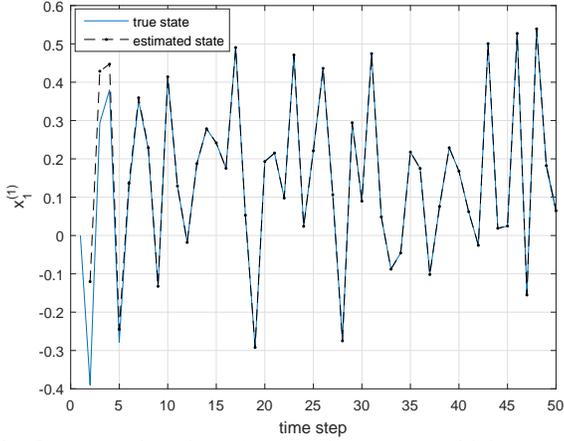}\\
\vspace{-4mm}
  \caption{The estimation of the MP state by utilizing the filter (\ref{eq:x1-reconstruction-state}).}\label{fig:x1}
\vspace{-3mm}
\end{figure}

\begin{figure}
  \centering
  \includegraphics[width=0.47\textwidth]{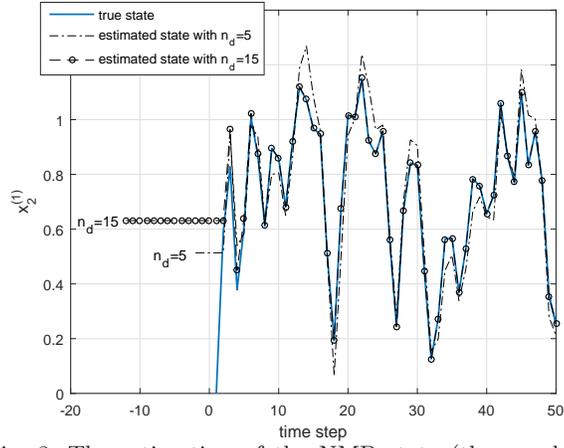}\\
\vspace{-4mm}
  \caption{The estimation of the NMP state (the graphs are shifted by $n_d-n$ time steps to the left for the purpose of comparison) by utilizing the filter (\ref{eq:nmp-state-reconstruction}).}\label{fig:x2}
\vspace{-2mm}
\end{figure}

\begin{figure}
  \centering
  \includegraphics[width=0.47\textwidth]{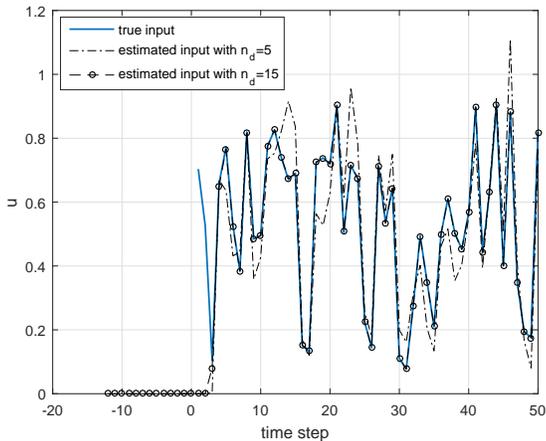}\\
\vspace{-4mm}
  \caption{The estimation of the unknown input (the graphs are shifted by $n_d-n$ time steps to the left for the purpose of comparison) by utilizing equation (\ref{eq:uhat-d}).}\label{fig:u}
\vspace{-2mm}
\end{figure}

\textbf{Case I:} Consider the following discrete-time linear system,
\begin{equation}
G(z)=\frac{(z-1.5)(z-0.5)}{z^2}
\end{equation}
or in its equivalent state space representation given by,
\begin{equation}
\left\lbrace\begin{array}{l} x(k+1)=\left[\begin{array}{cc}0 &0 \\1 &0\end{array}\right]x(k)+\left[\begin{array}{c}0 \\1 \end{array}\right]u(k) \\
y(k)=\left[\begin{array}{cc}-2& 0.75\end{array}\right]x(k)+u(k)
\end{array}\right.
\end{equation}
Using Lemma \ref{lm:sylvester solution}, the solution to the conditions (\textbf{i})-(\textbf{iii})  is given by $\hat{A}=0.5$, $F=\left[\begin{array}{cc}-0.5547& 0\end{array}\right]$, and $\mathbf{M}=\left[\begin{array}{cc}-0.5574& 0.8321\end{array}\right]$. Therefore, the unknown input observer is now given by equation (\ref{eq:x1-reconstruction-state}),
\begin{equation}
\left\lbrace \begin{array}{l}\eta(k+1)=0.5\eta(k)-0.5574y(k) \\ \hat{x}_1^{(1)}(k)=\eta(k)
\end{array}\right.
\end{equation}
Moreover, we have from the LQ decomposition of $\mathbf{M}$ and equations (\ref{eq:azd}) and (\ref{eq:bzd}), the following
\[\mathbf{T}^{(1)}=\left[\begin{array}{cc}-0.5547  &  0.8321\\0.8321  &  0.5547\end{array}\right];L=\left[\begin{array}{cc}1 &0 \end{array}\right]\]
\[A_{zd}=1.5;B_{zd}=\left[\begin{array}{cc}-1.5 &-1 \end{array}\right].\]
The upper bound for the state estimation error versus $n_d$ is shown in Figure \ref{fig:nd}. We have applied a \underline{non-smooth random input} to the system in order to illustrate and demonstrate the effects of $n_d$ on the estimation error. A smooth input, as stated in Remark \ref{rm:small-nd}, will be estimated in an almost unbiased manner for any $n_d \geq 2$. Figure \ref{fig:x1} depicts that $x_1^{(1)}(1)$ is perfectly estimated by using the unknown input observer (UIO) as expected. Figure \ref{fig:x2} shows that a perfect reconstruction can be achieved for $x_2^{(1)}(1)$ by selecting $n_d=15$,  as expected from Figure \ref{fig:nd}. According to the Proposition \ref{prop:input-reconstruction error-B1}, the unknown input should also be almost perfectly reconstructed with $n_d=15$, which is also verified in Figure \ref{fig:u}.

\begin{figure}
  \centering
  \includegraphics[width=0.47\textwidth]{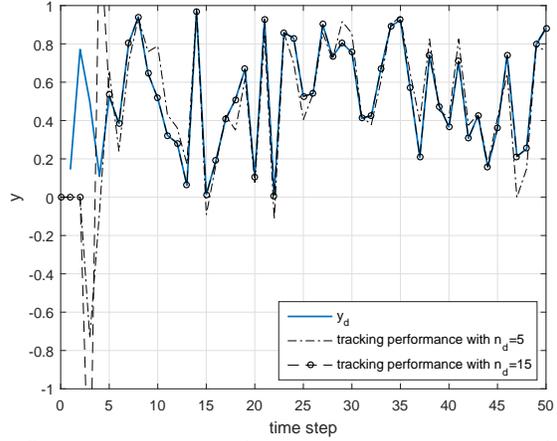}\\
\vspace{-4mm}
  \caption{The output tracking performance corresponding to different values of $n_d$ by utilizing equations (\ref{eq:outputTracking-x1}), (\ref{eq:outputTracking-x2}) and (\ref{eq:outputTracking-uhat-d}).}\label{fig:y}
\vspace{-3mm}
\end{figure}

\vspace{-3mm}
\textbf{Case II:} In another simulation case study, consider that a non-smooth $y_d(k)$ is required to be followed. The unknown input is reconstructed by using the  Algorithm that is detailed in Table \ref{tab:output-tracking-algorithm}, and the results are depicted in Figure \ref{fig:y}. This figure demonstrates that an almost perfect output tracking is achieved by selecting $n_d=15$. Finally, consider a smooth $y_d(k)$ as given by $y_d(k)=k^2\sin(5\pi k)$.  {The output tracking result for this smooth desired trajectory (not shown due to space limitations) confirms and validates the statements  made in Remark \ref{rm:small-nd}}.
\vspace{-5mm}

\vspace{1.5mm}
\textbf{Case III:} To provide a comparative study, consider a MIMO system that is taken from the reference \cite{Marro2010815} with $A \in \mathbb{R}^{4 \times 4}$, $B \in \mathbb{R}^{4 \times 2}$ and $C \in \mathbb{R}^{2 \times 4}$ as follows,
\begin{equation}\label{eq:zatt-sys}
\scriptsize{
\left\lbrace\begin{array}{l} x(k+1)=\left[\begin{array}{cccc}0.6 &-0.3& 0&0 \\0.1 &1 &0 &0\\ -0.4& -1.5& 0.4& -0.3\\ 0.3 &1.1& 0.2&0.9\end{array}\right]x(k)+\left[\begin{array}{cc}0 &0.4\\0&0 \\ 0&-0.1 \\0.1 &0.1 \end{array}\right]u(k) \\
y(k)=\left[\begin{array}{cccc}1& 2& 3&4\\ 2& 1& 5& 6\end{array}\right]x(k)
\end{array}\right.}
\end{equation}
The system (\ref{eq:zatt-sys}) has two zeros at $z_1=0.6072$ and $z_1=1.9928$. Therefore, it has three MP states and one NMP state. Authors of \cite{Marro2010815} proposed a geometric approach and applied it to the system (\ref{eq:zatt-sys}) to achieve an \textit{almost} perfect estimation of the states and unknown inputs with a delay of 20 time steps ($n_d=20$). For comparison, our simulation results for the same example is shown in Figure \ref{fig:zatt} (the numerical values of the estimation filter parameters are given in Appendix \ref{app:zatt}), which demonstrate that by using our proposed methodology the unknown states and inputs are almost perfectly reconstructed with only a delay of $n_d=10$, which is half of the delay that was used in \cite{Marro2010815}. Moreover, as shown in Figure \ref{fig:zatt}b, by using our approach the three MP states of the system are estimated without any delay when the transient response due to the unknown initial  condition dies out quick!
 ly. This is in contrast to the delayed results that are shown in the work \cite{Marro2010815}.\\

\vspace{-3mm}
{However, the \underline{most} important contribution of our work over that in \cite{Marro2010815} is derived from the fact that our methodology unlike the one in \cite{Marro2010815} \underline{can} handle transmission zeros on the unit circle as illustrated in the next case study.}\\

\vspace{-4mm}
{
\textbf{Case IV:} Consider the following system,
\begin{equation}\label{sys:unitcircle}
G(z)=\frac{(z+1)(z+3)(z+0.5)(z-0.5)}{z^2(z^2-z+0.5)}
\end{equation}
The above system has both MP and NMP transmission zeros as well as a zero on the unit circle. We follow the procedure that was introduced in Section \ref{sec:unitcircle} for designing an inversion-based output tracking controller. We selected the controller $H(z)$ having the structure,
\begin{equation}\label{eq:hz-z1}
H(z)=\frac{z+1}{2z}\mathbf{I}
\end{equation}
The numerical values for the other parameters are given in Appendix \ref{app:unitcircle}. Figure \ref{fig:unitcircle} shows the output tracking performance of our proposed solution with $n_c=1$ and $n_d=10$. The result demonstrates the significant advantage of our proposed solution for handling all types of transmission zeros within a \underline{single} framework. Specifically, Figure \ref{fig:unitcircle} shows that the desired trajectory is approximately followed by an error that is governed by equation (\ref{eq:error-uhz}). On the other hand, the proposed methodology in \cite{Marro2010815} essentially fails under this case. Note that the $H(z)$ that is selected in equation (\ref{eq:hz-z1}) can be used for all SISO systems that have a transmission zero at $z=-1$, in addition to MP and NMP zeros.}

\begin{figure}
  \centering
  \begin{tabular}{cc}
    a \\
    \includegraphics[width=0.41\textwidth]{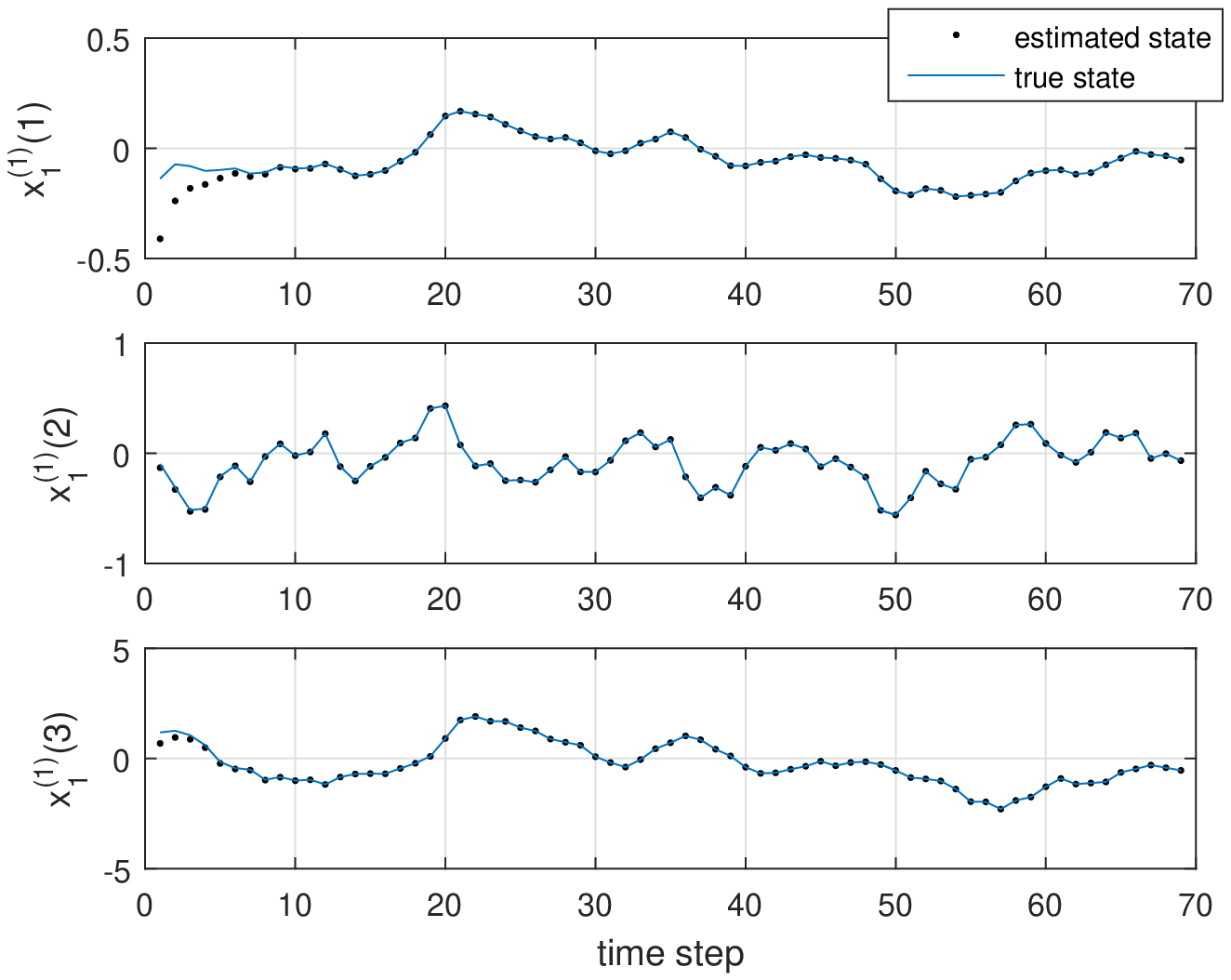}\\b \\
       \includegraphics[width=0.41\textwidth]{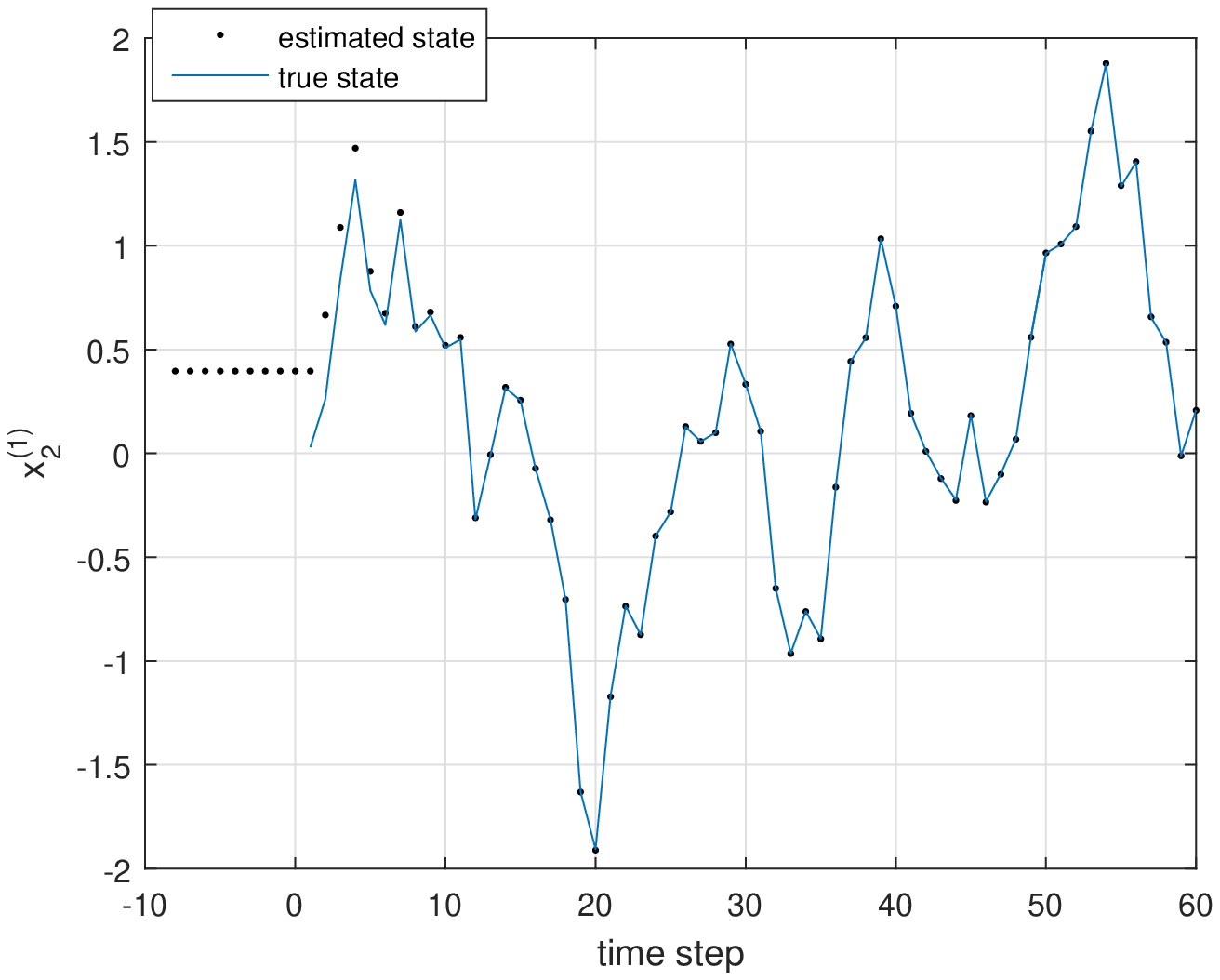} \\ c\\
      \includegraphics[width=0.41\textwidth]{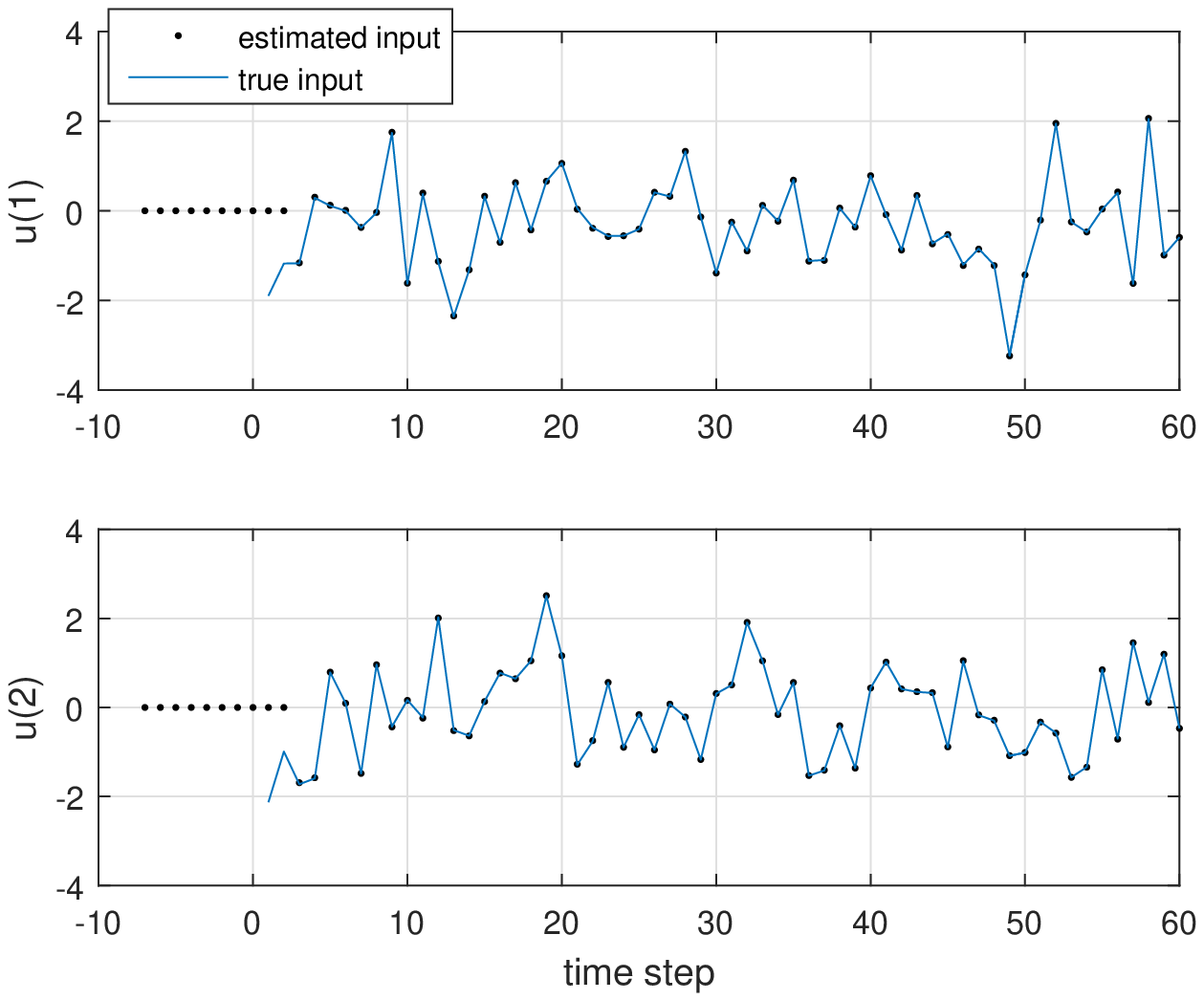} \\
  \end{tabular}
\vspace{-4mm}
    \caption{ Simulation results for the MIMO system (\ref{eq:zatt-sys}) taken from \cite{Marro2010815}, (a) The MP state estimates, (b) The NMP state estimates, and (c) The unknown input estimates.}\label{fig:zatt}
\vspace{-1.5mm}
\end{figure}
\begin{figure}
  \centering
  \includegraphics[width=0.47\textwidth]{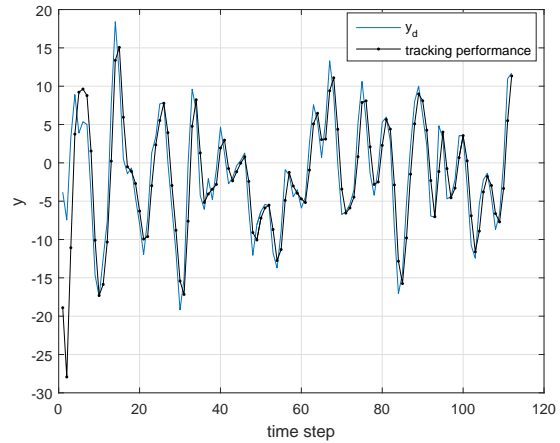}\\
  \caption{The output tracking performance for the system (\ref{sys:unitcircle}).}\label{fig:unitcircle}
\end{figure}
\vspace{-4mm}
\section{Conclusion}
\vspace{-4mm}
In this paper, we have shown that one can almost perfectly estimate and reconstruct the unknown state and inputs of a system if i) the system $\mathbf{S}$ is square, and ii) $B_1^{(1)}$ or $D$ is full column rank. Non-square systems rarely have transmission zeros \cite{DAVISON1974643}, and therefore it is straightforward to design an unknown input observer (UIO) to estimate all the system states. We excluded non-square systems from our analysis since Theorem \ref{thm:zeros-of-a-bipkc} is not guaranteed for this class of systems. In other words, the eigenvalues of $\Gamma=(A-B\mathbf{I}_n\mathbf{D}_n^+\mathbf{C}_n)$ may or may not coincide with the transmission zeros of the system. Also, it may or may not have the same characteristics, namely the MP transmission zero of $\mathbf{S}$ remains the stable eigenvalue of $\Gamma$. However, if one determines the matrices $\hat{A}$, $F$ and $\mathbf{M}$ by using a different method for these systems, then the remainder of our procedur!
 e for unknown state and input reconstruction, as described in this paper, will remain applicable and unchanged. We have also demonstrated that our proposed methods can provide an almost perfect tracking of any desired output trajectory by using data and information that correspond to a small preview time. {An important contribution of our methodology is the fact that we have provided a \underline{single} framework that can handle the problem of output tracking for systems that have transmission zeros on the unit circle in addition to MP and NMP zeros.} However, further research is required to address  issues of robustness and tracking error performance in presence of disturbances and modeling uncertainties.   These issues are left as topics of future research.

\bibliographystyle{ieeetr}
\bibliography{engine}

\appendix
\vspace{-2mm}
\section{Proof of Theorem \ref{thm:zeros-of-a-bipkc}}\label{app:zeros-of-a-bipkc}
The eigenvalues of $A-B\mathbf{I}_n\mathbf{D}_{n}^+\mathbf{C}_{n}$ are obtained by solving $|z\mathbf{I}-A+B\mathbf{I}_n\mathbf{D}_{n}^+\mathbf{C}_{n}|=0$. If the system is square, then $\mathbf{D}_{n}^+$ is a nonzero square matrix. Therefore, one can equivalently solve the equation $\left|\mathbf{D}_{n}^+\right||z\mathbf{I}-A+B\mathbf{I}_n\mathbf{D}_{n}^+\mathbf{C}_{n}|=0$. On the other hand from the Schur identity { \cite{schur}}, we have,
\begin{equation}\label{eq:schur}
\left|\mathbf{D}_{n}^+\right||z\mathbf{I}-A+B\mathbf{I}_n\mathbf{D}_{n}^+\mathbf{C}_{n}| = \left|\left[\begin{array}{cc}z\mathbf{I}-A &-B\mathbf{I}_n \\ \mathbf{C}_{n} & \mathbf{D}_{n} \end{array}\right]\right|
\end{equation}
Let us partition $\mathbf{C}_{n}$ and  $\mathbf{D}_{n}$ as follows,
\begin{equation}
\mathbf{C}_{n} = \left(\begin{array}{c} C \\ \hline \mathcal{C}^- \end{array} \right)=\left(\begin{array}{c}C\\ \hline CA\\ \vdots \\ CA^{n-1}\end{array}\right)
\end{equation}
\begin{equation}
\mathbf{D}_{n}=\left(\begin{array}{c|c}D & 0 \\ \hline \mathcal{D}^-_{21} & \mathcal{D}^-_{22}\end{array}\right) =\left(\begin{array}{c|ccc} D&0& \ldots & 0\\ \hline CB& D& \ldots &0 \\ \vdots & \vdots & \vdots & \vdots \\ CA^{n-1}B&CA^{n-2}B&\ldots & D\end{array}\right)
\end{equation}
Then, the right hand side of equation (\ref{eq:schur}) can be partitioned as,
\begin{equation}
\left[\begin{array}{cc}
z\mathbf{I}-A &-B\mathbf{I}_p \\ \mathbf{C}_{n} & \mathbf{D}_{n} \end{array}\right]= \left[\begin{array}{cc|c}
z\mathbf{I}-A &-B & 0\\ C & D & 0 \\ \hline \mathcal{C}^- & \mathcal{D}^-_{21} &\mathcal{D}^-_{22} \end{array}\right]
\end{equation}
Thus, if $\mathcal{D}^-_{22}$ is full row rank, then according to the Schur identity, equation $|z\mathbf{I}-A+B\mathbf{I}_n\mathbf{D}_{n}^+\mathbf{C}_{n}|=0$ has only one set of solution that is given by $\left|\left[\begin{array}{cc}
z\mathbf{I}-A &-B\mathbf{I}_p \\ \mathbf{C}_{n} & \mathbf{D}_{n} \end{array}\right]\right|=0$, which is exactly the transmission zeros of the system $\mathbf{S}$. However, if $\mathcal{D}^-_{22}$ is rank deficient, then certain rows of $\left[\begin{array}{ccc}\mathcal{C}^- & \mathcal{D}^-_{21} &\mathcal{D}^-_{22}\end{array}\right]$ are linearly dependent on the rows of $\left[\begin{array}{ccc}-A& -B &0\end{array}\right]$. Hence, $z=0$ is also a solution. On the other hand, since $|z\mathbf{I}-A+B\mathbf{I}_n\mathbf{D}_{n}^+\mathbf{C}_{n}|=0$ must have $n$ eigenvalues, if the system $\mathbf{S}$ has $p$ transmission zeros, then $z=0$ is a solution of multiplicity $n-p$.  $\blacksquare$
\vspace{-3mm}

\vspace{-3mm}
\section{Proof of Theorem \ref{thm:complete-solution}}\label{app:complete-solution}
Since the system $\mathbf{S}$ has $\alpha_1$ transmission zeros having an algebraic multiplicity of 1, therefore $\Gamma$ has $\alpha_1$ linearly independent eigenvectors. Therefore, $\mathbf{M}_0$ has at least $\alpha_1$ linearly independent rows. On the other hand, the set $\mathcal{Z}$ (as defined in Theorem \ref{thm:zeros-of-a-bipkc}) has $\alpha_z$ zeros, where $\alpha_z=n-\beta-\alpha_1$.  Therefore, $\mathbf{I}-\mathbf{D}_n\mathbf{D}_n^+$ has $\alpha_z$ independent rows.  This implies that $\mathbf{M}_\#$ has $\alpha_z$ linearly independent rows. Therefore, $\mathbf{M}$ has $\alpha_z+\alpha_1=n-\beta$ linearly independent rows.  $\blacksquare$
\vspace{-4mm}
\section{Proof of Lemma \ref{lm:linear-dependence}}\label{app:linear-dependence}
Since the system $\mathbf{S}$ has at least one NMP zero ($q<n$), then by the definition of transmission zeros, there exists a nonzero $u(k)$ that yields a zero output ($y(k)=0$ for all $k$). On the other hand, according to Theorem \ref{thm:x1-feature}, $x_1^{(1)}(k)$ approaches to zero when $y(k)=0$ for  $k=k_0, k_0+1, \ldots$. Therefore, from the first and  third equations of (\ref{eq:system-partition}), we have for $k \rightarrow \infty$,
\begin{equation}
\left[\begin{array}{c}0\\ 0\end{array}\right]= \left[\begin{array}{cc}C_2^{(1)}&D \\ A_{12}^{(1)}& B_1^{(1)}\end{array}\right]\left[\begin{array}{c}x_2^{(1)}(k-n)\\u(k-n)\end{array}\right].
\end{equation}
Since $\left[\begin{array}{c}x_2^{(1)}(k-n)\\u(k-n)\end{array}\right]$ is nonzero, it implies that the columns of $\left[\begin{array}{cc}C_2^{(1)}&D \\ A_{12}^{(1)}& B_1^{(1)}\end{array}\right]$ are linearly dependent.  $\blacksquare$
\vspace{-4mm}
\section{Proof of Lemma \ref{lm:nmp-zeros-b}}\label{app:nmp-zeros-b}
First note that $A_{11}^{(1)}$ in equation (\ref{eq:system-partition}) is a Hurwitz matrix, otherwise $x_1^{(1)}(k) \rightarrow \infty$ as $k\rightarrow \infty$. Next,  consider,
\begin{equation}\label{eq:zero-dynamic-1}
S^{(z1)}:\left\lbrace\begin{array}{l} x_2^{(1)}(k-n+1)=A_{22}^{(1)}x_2^{(1)}(k-n)+B_2^{(1)}u(k-n)\\ \xi(k-n)=A_{12}^{(1)}x_2^{(1)}(k-n)+B_1^{(1)}u(k-n)\end{array}\right.
\end{equation}
If there exists a nonzero $u(k)$ that yields $\xi(k)=0$, then this implies that from the first  equation of (\ref{eq:system-partition}) we have, $x_1^{(1)}(k)\rightarrow 0$ as $k\rightarrow \infty$. Therefore, $y(k)\rightarrow 0$ as $k\rightarrow \infty$ according to Theorem \ref{thm:x1-feature}. Therefore, the transmission zeros of $S^{(z1)}$ are also the transmission zeros of $S^{(1)}$.  $\blacksquare$
\vspace{-4mm}
\section{Proof of Lemma \ref{lm:nmp-zeros-d}}\label{app:nmp-zeros-d}
Consider the following system,
\begin{equation}\label{eq:zero-dynamic-2}
S^{(z2)}:\left\lbrace\begin{array}{l} x_2^{(1)}(k-n+1)=A_{22}^{(1)}x_2^{(1)}(k-n)+B_2^{(1)}u(k-n)\\ \xi(k-n)=\left[\begin{array}{cc}C_2^{(1)}&D\end{array}\right]\left[\begin{array}{c}x_2^{(1)}(k-n)\\u(k-n)\end{array}\right]\end{array}\right.
\end{equation}
If there exists a nonzero $u(k)$ that yields $\xi(k)=0$, then since $\left[\begin{array}{cc}\mathbf{I} & C_1\end{array}\right]$ is full row rank, the third equation of (\ref{eq:system-partition}) yields $x_1^{(1)}(k)=0$, and $y(k)=0$. Therefore, the transmission zeros of $S^{(z2)}$ are also the transmission zeros of $S^{(1)}$. $\blacksquare$
\vspace{-4mm}
\section{Proof of Theorem \ref{thm:sz-properties}}\label{app:sz-properties}
Note that $C_{z2}=0$ is an immediate result of the Schur identity {\cite{schur}} and Lemma \ref{lm:linear-dependence}.  The eigenvalues of $A_z$  are a subset of the transmission zeros of $\left[\begin{array}{cc}A_{22}^{(1)} & B^{(1)}_2 \\ A_{12}^{(1)} & B_1^{(1)}\end{array}\right]$ , which are a subset of the system $\mathbf{S}$ zeros according to Lemma \ref{lm:nmp-zeros-b}. According to Theorem \ref{thm:x1-feature} and Theorem \ref{thm:sz-properties} ($C_{z2}=0$), the output of the system (\ref{eq:system-sz}) goes to zero as $k\rightarrow \infty$ if and only if $x_1^{(1)}(k)$, and consequently, $X_1^{(1)}(k)$  goes to zero as $k\rightarrow\infty$. The first equation of (\ref{eq:system-sz}) implies that if $A_z$ is a Hurwitz matrix, then $x_2^{(1)}(k)$ must approach to zero when $X_1^{(1)}(k)$ is zero.  However, we know that there exists  nonzero $x_2^{(1)}(k)$ and $u(k)$ that yield a zero $y(k)$ for all $k$. Therefore, since the response of an unforced line!
 ar system can approach to zero or infinity (recall we excluded systems with transmission zeros on the unit circle in Assumption 1), therefore  $x_2^{(1)}(k)$ must approach to infinity.  This implies that the eigenvalues of $A_z$ are the NMP zeros of $\mathbf{S}$.  $\blacksquare$
\vspace{-4mm}
\section{Proof of Theorem \ref{thm:sz-properties-d}}\label{app:sz-properties-d}
 e NMP zeros of $\mathbf{S}$.
{Proof follows along the same lines as those for Theorem \ref{thm:sz-properties} using Lemma \ref{lm:nmp-zeros-d} and is therefore omitted for brevity}.
$\blacksquare$
\vspace{-4mm}
\section{Numerical values for estimation filter parameters for the system (\ref{eq:zatt-sys})}\label{app:zatt}
\vspace{-4mm}
\[\hat{A}=\left[\begin{array}{ccc} 0.6072 &0&0\\0&0&0\\0&0&0\end{array}\right], ~~A_z=1.9928\]
\[F=\left[\begin{array}{cccc} 0&0&-0.662&0.0184\\0&0&-0.0206&0.1365\\0&0&0.1337&0.0338 \end{array}\right]\]
\[\mathbf{M}=\left[\begin{array}{cccc} 0.0488 &   0.9650  &  0.2063  & -0.1547\\0.2523  &  0.0953  &  0.6205 &   0.7364\\0.2013  &  0.3012  &  0.5700  &  0.7375 \end{array}\right]\]
\[\mathbf{T}^{(1)}=\left[\begin{array}{cccc} -0.0488  & -0.9650 &  -0.2063 &   0.1547\\ 0.2483 &  -0.0190  &  0.6003&    0.7600\\-0.4645  &  0.2474 &  -0.5833 &   0.6187\\  -0.8487  & -0.0855   & 0.5067 &  -0.1252 \end{array}\right]\]
\[L=\left[\begin{array}{cccc} -1.0000   &      0    &     0     &    0\\ -0.1183  &  0.9930   &      0   &      0\\-0.3039  &  0.9469   & 0.1048      &   0 \end{array}\right]\]
\[B_z=\left[\begin{array}{cccccc} -0.2463  & -2.0822  &  2.4171  &  5.3035  & -0.0678 &  -2.3507 \end{array}\right]\]
\vspace{-5mm}

{
\vspace{-5mm}
\section{Numerical values for estimation filter parameters for the system (\ref{sys:unitcircle})}\label{app:unitcircle}
\vspace{-4mm}
\[\hat{A}=\left[\begin{array}{ccc} 0 &0&0\\0&0.5&0\\0&0&-0.5\end{array}\right],F=\left[\begin{array}{cccc} 0&1.042&0&0\\0&-0.4119&0&0\\0&0.4823&0&0 \end{array}\right]\]
\[\mathbf{M}=\left[\begin{array}{cccc} 0.368 &   -0.368  &  0.792  & -0.317\\0  &  0.395  &  -0.852 &   0.341\\0  &  -0.395  &  -0.852 &   0.341 \end{array}\right]\]
\[\mathbf{T}^{(1)}=\left[\begin{array}{cccc} -0.3680  &  0.3680 &  -0.7928  &  0.3171\\
   -0.9298 &  -0.1456 &   0.3138 &  -0.1255\\
    0 &  -0.9184 &  -0.3674 &   0.1470\\
   0 &  0 &   0.3714   & 0.9285 \end{array}\right]\]
\[L=\left[\begin{array}{cccc} -1.0000     &    0  &       0   &      0
 \\   0.9298 &  -0.3680      &   0      &   0\\
    0.6386  & -0.2527   & 0.7269   &      0\end{array}\right], A_z=3\]
\[B_z=\left[\begin{array}{cccccc}     2.1348  & -0.8448   & 0.9894 &  -0.2135 &   0.0845&   -0.0989
 \end{array}\right]\]}

\end{document}